\documentclass[amsmath,amssymb,amsfonts,twocolumn]{revtex4-1}
\usepackage{bbm}
\usepackage{braket,stmaryrd,epsfig}
\usepackage{mathtools,graphicx}
\def \beq {\begin{equation}}
\def \eeq {\end{equation}}
\def \tr {\rm Tr}
\newcommand {\qs} {{\rm Q}_{\rm S}}
\newcommand {\qt} {{\rm Q}_{\rm T}}
\newcommand {\ks} {k_{\rm S}}
\newcommand {\kt} {k_{\rm T}}
\newcommand{\bomb}{
 {\mathchoice
  {\includegraphics[height=1.6ex]{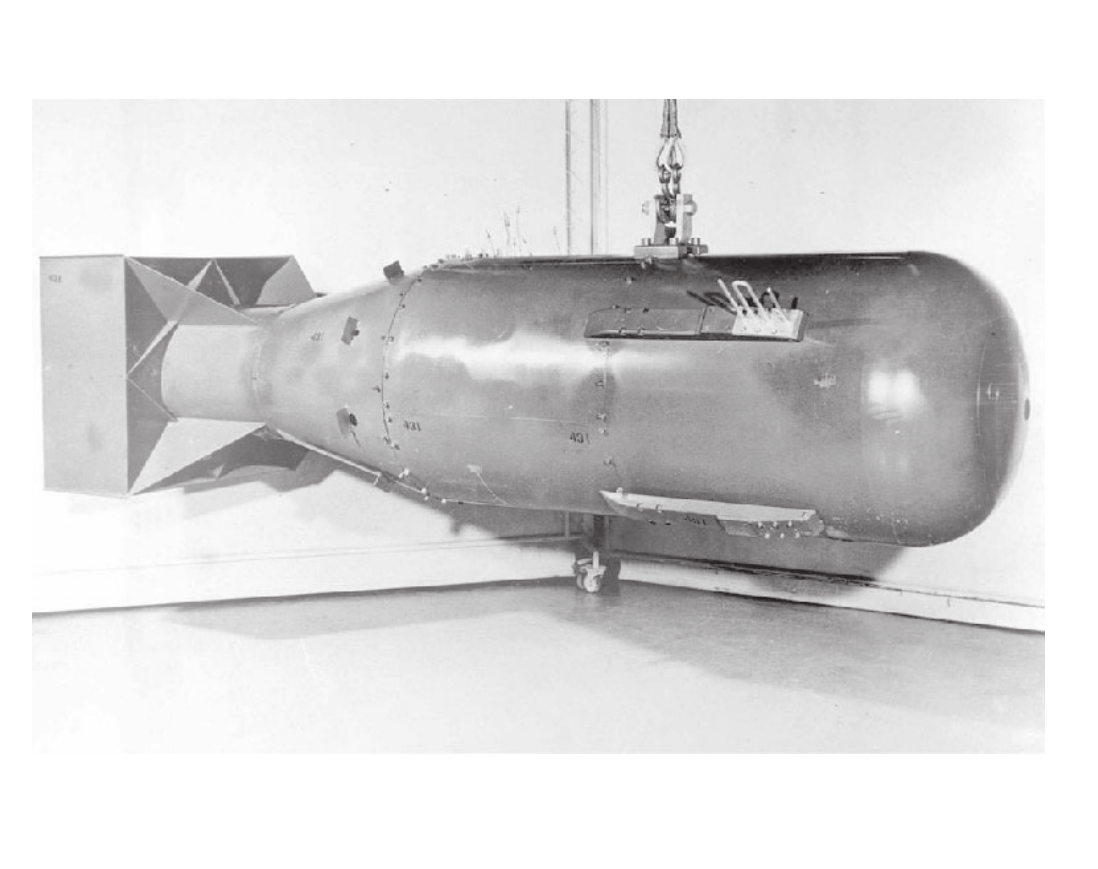}}
  {\includegraphics[height=1.6ex]{bomb.pdf}}
  {\includegraphics[height=1.2ex]{bomb.pdf}}
  {\includegraphics[height=0.9ex]{bomb.pdf}}
 }
}
\newcommand{\explosion}{
 {\mathchoice
  {\includegraphics[height=1.6ex]{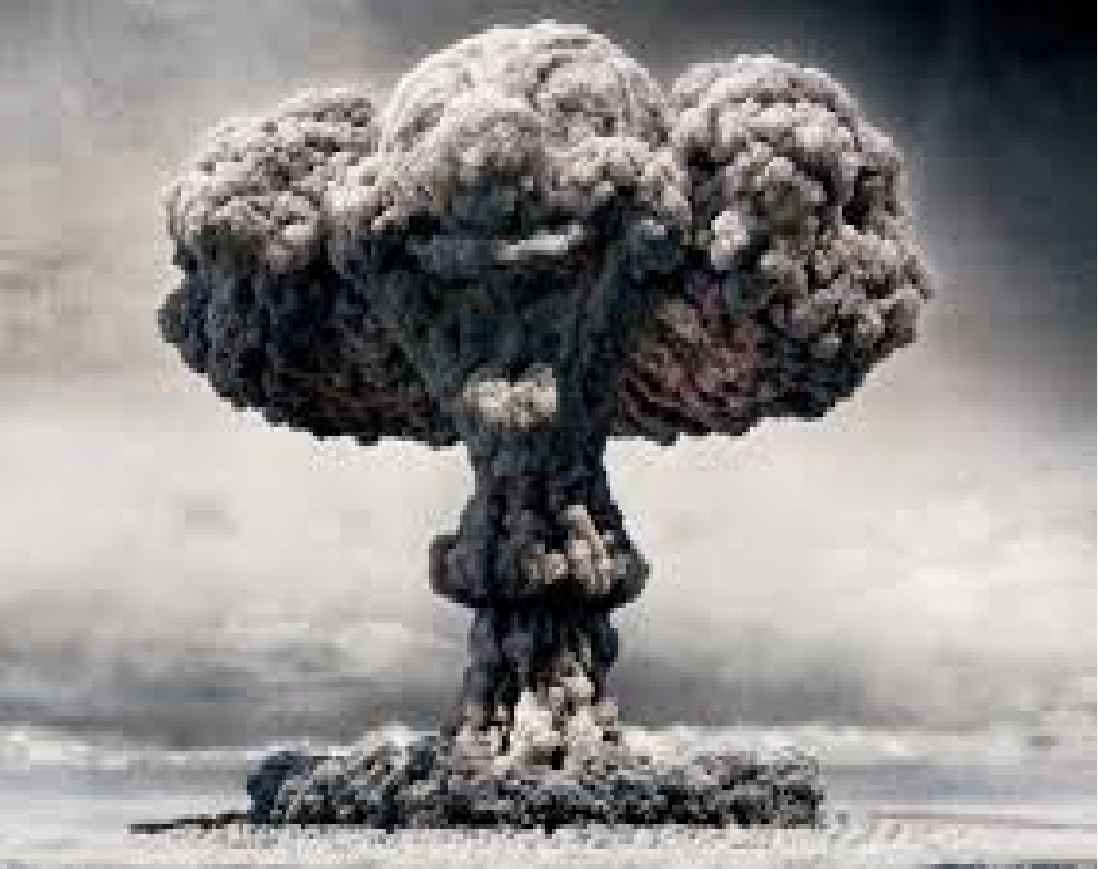}}
  {\includegraphics[height=1.6ex]{explosion.pdf}}
  {\includegraphics[height=1.2ex]{explosion.pdf}}
  {\includegraphics[height=0.9ex]{explosion.pdf}}
 }
}

\begin{document}

\title{The radical-pair mechanism as a paradigm for the emerging science of \\quantum biology}
\author{Iannis K. Kominis}

\affiliation{Department of Physics, University of Crete, Heraklion 71103, Greece}

\begin{abstract}
The radical-pair mechanism was introduced in the 1960's to explain anomalously large EPR and NMR signals in chemical reactions of organic molecules. It has evolved to the cornerstone of spin chemistry, the study of the effect electron and nuclear spins have on chemical reactions, with the avian magnetic compass mechanism and the photosynthetic reaction center dynamics being prominent biophysical manifestations of such effects. In recent years the radical-pair mechanism was shown to be an ideal biological system where the conceptual tools of quantum information science can be fruitfully applied. We will here review recent work making the case that the radical-pair mechanism is indeed a major driving force of the emerging field of quantum biology.
\end{abstract}

\maketitle

\section{Introduction}
Quantum biology is swiftly emerging as an interdisciplinary science synthesizing, in novel and unexpected ways, the richness and complexity of biological systems with quantum physics and quantum information science. 
The intuitive disposition prevailing until recently, namely that quantum coherence phenomena have no relevance to biology, can be simply understood. Indeed, quantum coherence has been clearly manifested in carefully prepared and well-isolated quantum systems, the experimental implementation of which was technically demanding. It was thus plausible to assume that all timescales of interest in complex biological matter 
are orders of magnitude larger than the coherence time of any underlying quantum phenomenon. In other words, decoherence has been assumed to be detrimental merely due to the complexity of biological systems and the vast environments they offer as sinks of quantum information, environments that were understood to be anything but "carefully prepared".

In recent years, however, a different physical picture is gradually emerging. Biological systems are indeed nothing like single atoms trapped in ultra-high vacuum, allowing the observation of coherences lasting for several seconds \cite{deutsch} or even hours for solid-state nuclear spin systems \cite{sellars}. But they are also nothing like a classical system where decoherence has obliterated any chance to detect off-diagonal elements of the relevant density matrix \cite{schlosshauer}. It appears that nature has optimized biological function using both worlds, i.e. taking advantage of quantum coherence amidst an equally essential yet not detrimental decoherence. 

In this review we focus on a major driving force of quantum biology, namely the quantum dynamics of the radical-pair mechanism (RPM).  The RPM involves a multi-spin system of electrons and nuclei embedded in a biomolecule and undergoing a number of physical/chemical processes, like magnetic interactions and coherent spin motion, electron transfer and spin relaxation to name just a few. For an overview of RPM's long scientific history we refer the reader to a number of reviews of the earlier work on radical-ion pairs and spin chemistry \cite{steiner1,steiner2,woodward,rodgers_review}. 

Although the RPM has been known since the 1960s, the rich quantum physical underpinnings of the mechanism have been unraveled only recently. The first paper \cite{pre2009} discussing RPM in the context of modern quantum information theory introduced quantum measurements as a necessary concept to understand the  quantum dynamics of RPM, leading among other things to the fundamental spin decoherence process of radical-ion-pair (RP) reactions. In a number of papers that followed \cite{njp2010,cpl2011,pre2011,pre2012,cpl2012_1,cpl2012_2,biosys,cidnp,lamb,pre2014,cpl2015}, the intricate quantum effects at play in RP reactions were further elucidated. Quantum measurement dynamics, phononics, in particular the phonon vacuum, quantum coherence and decoherence, in particular measurement-induced decoherence, coherence distillation, quantum coherence quantifiers, quantum correlations, quantum jumps and the quantum Zeno effect, concepts from quantum metrology and the quantum-communications concept of quantum retrodiction were shown to be central for understanding RPM.

Were it not for these conceptual underpinnings, it would be hard to justify why RPM constitutes a paradigm for quantum biology. It can of course be argued that radical-ion pairs involve spins, and spins are genuine quantum objects, but this argument points to the atomistic, structural aspect of biology. That is, one could also argue that spins, e.g. singlet states in the helium atom, determine atomic structure and atomic structure is the basis of biological structure, alas this is not what quantum biology is about. We hope, however, to convince the reader that RPM's position on the map of quantum biology is indeed well-deserved. 

This review will focus on our recent work concerning the quantum foundations of RPM. This biochemical spin system is an open system and hence the relevant density matrix $\rho$ does not just obey Schr\"{o}dinger's equation, $d\rho/dt=-i[{\cal H},\rho]$, where ${\cal H}$ is the system's Hamiltonian, but more terms are required to account for the "open" character of RPs. These terms are compactly written as a "reaction super-operator" ${\cal L}(\rho)$. The master equation $d\rho/dt=-i[{\cal H},\rho]+{\cal L}(\rho)$ is the starting point for all theoretical calculations involving RP reactions, hence the foundational character of this discussion. Our specific quest has been to find the exact form of the reaction super-operator ${\cal L}(\rho)$. Until we entered this field in 2008, the overwhelming majority of theoretical treatments in spin chemistry have used what we now understand is a phenomenological theory elaborated upon by Haberkorn \cite{haberkorn} in 1976. 

Returning to the chronological exposition, the first paper \cite{pre2009} illuminating the quantum-information substratum of RPM led to a resurgent interest of the spin chemistry community on the foundational aspects of the field \cite{shushin,purtov,ivanov,maeda2011,molin2013,molin2013_2,maeda2013,purtov2,berdinskii,manolopoulos}, and in parallel to a flurry of activity in the quantum science community\cite{JH,briegel,vedral,cai,JHreply,plenioPRA,sun1,sun2,stoneham,horePRL,kaszlikowski,tiersch,briegelJPC,plenioPRL,shao,comment1,comment2,kais1,tiersch2014,briegel2014,kais2,walters,shao2014,whaley,poonia,kais3,oliveira}, even prompting some authors to announce, more or less timely, "the dawn of quantum biology" \cite{ball}. The main motivation and excitement behind this work has largely been RPM's relation to the avian magnetic compass. It is clear that concepts of quantum measurements, quantum (de)coherence and entanglement take on a distinct scientific flavor when applied to migrating birds. In this review we will only cover the part of the literature relevant to our specific quest of obtaining the master equation for $d\rho/dt$. We feel that the rest of the literature, especially papers discussing quantum coherence and entanglement in chemical magnetoreception, address a very promising research venue, however they are mostly based on the traditional description of RP dynamics. Since the qualitative and quantitative understanding of quantum coherence and entanglement is intimately tied to the master equation for $d\rho/dt$, these discussions could be premature.

Focusing on the RPM, we will not cover other, equally exciting venues of quantum biology, like quantum effects in photosynthetic light harvesting and olfaction recently reviewed \cite{plenio_review}, coupling quantum light to retinal rod cells \cite{kriv1,kriv2}, testing for electron spin effects in anesthesia \cite{turinAN}, and studying ion transport in neuronal ion channels \cite{vaziri1,vaziri2}, to name a few. 

The outline of this review is the following. In Section 2 we briefly review the basic notions of spin chemistry, touching upon the traditional approach to the radical-pair mechanism. In Section 3 we present our work on the quantum foundations of RPM. In light of this discussion, we revisit the traditional approach in Section 4 and explore its weaknesses and limits of validity. In Section 5 we study single-molecule quantum trajectories, which allow us to obtain a deeper understanding of both approaches and test their internal consistency. In Section 6 we discuss other competing theoretical approaches. We devote Section 7 to discussing the nonlinear nature of our master equation, and its relation to foundational concepts of quantum physics. We close with an outlook in Section 8.
\section{Spin Chemistry}
Spin chemistry \cite{hayashi,salikhov,molin_book} deals with the effect of electron and nuclear spins on chemical reactions. That such an effect is possible in the first place is not too straightforward to understand. Indeed, covalent and hydrogen bond energies are on the order of 1000 kJ/mole and 10 kJ/mole, respectively, while the electron's Zeeman energy in a laboratory magnetic field of 1 kG is on the order of 1 J/mole, let alone earth's field. Yet even at earth's field, electron and nuclear spins can have quite a tangible effect in this class of chemical reactions. The explanation, to be revisited later, is an intricate combination of spin precession and electron transfer timescales with spin angular momentum conservation.

Although the first organic free radical, triphenylmethyl, was discovered by Gomberg in 1900, the origin of spin chemistry is set in the 1960s, when anomalously high EPR \cite{cidep} and soon later NMR \cite{bargon,ward,cocivera} signals were observed in chemical reactions of organic molecules, termed CIDEP (chemically induced dynamic electron polarization) and CIDNP (chemically induced dynamic nuclear polarization), respectively. The radical-pair mechanism was introduced by Closs and Closs \cite{closs} and by Kaptein and Oosterhoff \cite{kaptein} as a reaction intermediate explaining these observations (a recent editorial  \cite{matysik_edit} briefly reviews the field's history and early literature). Since then, spin chemistry has grown into a mature field of experimental and theoretical physical chemistry \cite{steiner1}. In particular, the effects related to CIDNP \cite{beyerle,boxer,zys,mcdermott,vieth1,vieth2,vieth3,goez,prakash,diller,daviso,matysik_pnas,matysik_PR,jeschke_TSM,jeschke1998,JM,jeschke_lowB}, which we will address in Section 5.2, continue to attract a lot of interest since CIDNP has become a versatile tool to study photosynthetic reaction centers \cite{matysik_review}. 

An intriguing result of early spin-chemistry work was the proposal by Schulten \cite{schulten77,schulten1,schulten2} that avian magnetoreception is based on RP reactions. In this review we will not touch upon applying spin chemistry to magnetoreception, referring the reader to a representative part \cite{WWscience,ww78,munro,borland,sandberg,ritz2000,wild,WW,ritz2004,johnsen1,KS2007,johnsen2,horePNAS,ritz2009,KS2009,mour2009,mour2010,mour2012,mouritsen} of an extensive literature.
\subsection{Introduction to the radical-pair mechanism}
The quantum degrees of freedom of radical-ion pairs are formed by a multi-spin system embedded in a biomolecule, which can be either in the liquid or in the solid phase. In particular, RPs are biomolecular ions created by a charge transfer from a photo-excited D$^*$A donor-acceptor dyad DA, schematically described by the reaction ${\rm DA}\rightarrow {\rm D^{*}A}\rightarrow {\rm D}^{\bullet +}{\rm A}^{\bullet -}$, where the two dots represent the two unpaired electron spins of the two radicals. The excited state D$^*$A is usually a spin zero state, hence the initial spin state of the two unpaired electrons of the radical-pair is also singlet, denoted by $^{\rm S}{\rm D}^{\bullet +}{\rm A}^{\bullet -}$. Now, both D and A contain a number of  magnetic nuclei which hyperfine-couple to the respective electron. Neither singlet-state nor triplet state RPs are eigenstates of the resulting Hamiltonian, ${\cal H}$, hence the initial formation of $^{\rm S}{\rm D}^{\bullet +}{\rm A}^{\bullet -}$ is followed by singlet-triplet (S-T) mixing, i.e. a coherent oscillation of the spin state of the electrons, designated by $^{\rm S}{\rm D}^{\bullet +}{\rm A}^{\bullet -}\xleftrightharpoons{\cal H}~^{\rm T}{\rm D}^{\bullet +}{\rm A}^{\bullet -}$. Concomitantly, nuclear spins also precess, and hence the total electron/nuclear spin system undergoes a coherent spin motion driven by hyperfine couplings and the rest of the magnetic interactions to be detailed later. 

\begin{figure}
\begin{center}
\includegraphics[width=7 cm]{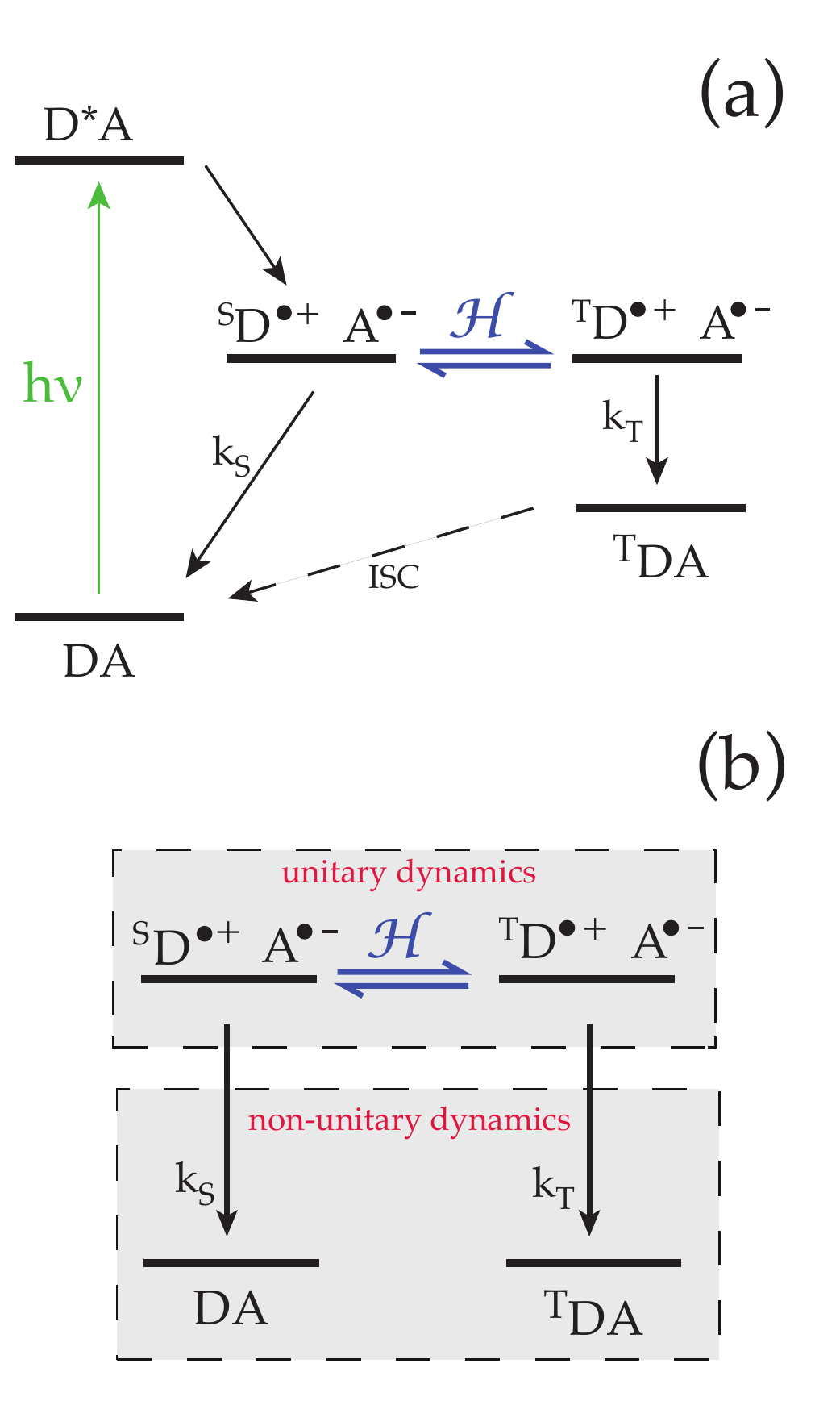}
\caption{(a) Radical-pair reaction dynamics. A donor-acceptor dyad DA is photoexcited to D$^*$A and a subsequent charge transfer produces a singlet state radical-pair $^{\rm S}{\rm D}^{\bullet +}{\rm A}^{\bullet -}$, which is coherently converted to the triplet radical-pair, $^{\rm T}{\rm D}^{\bullet +}{\rm A}^{\bullet -}$, due to intramolecule magnetic interactions.  Simultaneously, spin-selective charge recombination leads to singlet (DA) and triplet neutral products ($^{\rm T}$DA). The latter can intersystem cross into DA and close the reaction cycle. (b) Simplified version of (a) neglecting the photoexcitation and charge transfer steps. Both diagrams could be misleading if taken too literally, since they might suggest that e.g. only singlet radical-pairs recombine to singlet neutral products. This is not the case, since a radical-pair in a coherent S-T superposition can recombine into e.g. a singlet neutral product (see Section 3.7).}
\label{schematic}
\end{center}
\end{figure}

This coherent spin motion has, however, a finite lifetime. Charge recombination, i.e. charge transfer from A back to D, terminates the reaction and leads to the formation of the neutral reaction products. It is angular momentum conservation at this step that empowers the molecule's spin degrees of freedom and their minuscule energy to determine the reaction's fate: there are two kinds of neutral products, singlet (the original DA molecules) and triplet, $^{\rm T}$DA. As it turns out, their relative proportion can be substantially affected by spin interactions entering the mixing Hamiltonian ${\cal H}$. For completeness we note that the reaction can close through the so-called intersystem crossing $^{\rm T}{\rm DA}\rightarrow {\rm DA}$, mediated by e.g. spin-orbit coupling. A schematic diagram of the above is shown in Fig. \ref{schematic}a. 

Omitting the light excitation and initial charge transfer, which commence the reaction in a timescale usually faster than the reaction dynamics, the reaction scheme can be simplified into Fig. \ref{schematic}b. As shown with the two shaded boxes in Fig. \ref{schematic}b, the RP reaction consists of two physically very different processes working simultaneously: (a) the unitary dynamics embodied by the magnetic Hamiltonian ${\cal H}$ driving a coherent S-T oscillation of the RP spin state, and (b) the non-unitary reaction dynamics reducing the RP population in a spin-dependent way. As it will turn out, the latter are the hardest to understand.
\subsection{Definitions}
The density matrix $\rho$ describes the spin state of the RP's two electrons and $M$ magnetic nuclei located in D and A. Its dimension is $d=4\Pi_{j=1}^{M}(2I_{j}+1)$, where 4 is the spin multiplicity of the two unpaired electrons, $\Pi_{j=1}^{M}(2I_{j}+1)$ is the spin multiplicity of the $M$ nuclear spins, and $I_j$ is the nuclear spin of the $j$-th nucleus, with $j=1, 2,..., M$. The simplest possible RP contains just one spin-1/2 nucleus hyperfine coupled to e.g. the donor's unpaired electron. In this case the density matrix has dimension $d=8$. Although unrealistic, this simple system exhibits much of the essential physics without the additional computational burden of more nuclear spins and matrices of higher dimension, therefore it is frequently used as a model system.

It is angular momentum conservation at the recombination process that forces the decomposition of the RP's spin space into an electron singlet and an electron triplet subspace, defined by the respective projectors $\qs$ and $\qt$. These are $d\times d$ matrices given by $\qs={1\over 4}\mathbbmtt{1}_{d}-\mathbf{s}_{D}\cdot\mathbf{s}_{A}$ and $\qt={3\over 4}\mathbbmtt{1}_{d}+\mathbf{s}_{D}\cdot\mathbf{s}_{A}$, where $\mathbf{s}_{D}$ and $\mathbf{s}_{A}$ are the spin operators of the donor and acceptor electrons written as $d$-dimensional operators, e.g. the $j$-th component of $\mathbf{s}_{D}$ would be written as $s_{jD}=\hat{s}_{j}\otimes\mathbbmtt{1_2}\otimes\mathbbmtt{1}_{2I_1+1}\otimes\mathbbmtt{1}_{2I_2+1}...\otimes\mathbbmtt{1}_{2I_M+1}$, where the first operator in the Kronecker product refers to the donor's electron spin, the second to the acceptor's electron spin and the rest to the nuclear spins. By $\hat{\mathbf{s}}$ we denote the regular (2 dimensional) spin-1/2 operators and by $\mathbbmtt{1}_{m}$ the $m$-dimensional unit matrix. The projectors $\qs$ and $\qt$ are complete and orthogonal, i.e. $\qs+\qt=\mathbbmtt{1}_{d}$ and $\qs\qt=\qt\qs=0$.

The RP's singlet subspace has dimension $\Pi_{j=1}^{M}(2I_{j}+1)$ while the triplet subspace has dimension $3\Pi_{j=1}^{M}(2I_{j}+1)$. The electron-spin multiplicity 1 in the former corresponds to the singlet state 
\beq
\ket{\rm S}={1\over\sqrt{2}}\big(\ket{\uparrow\downarrow}-\ket{\downarrow\uparrow}\big),
\eeq
while the electron-spin multiplicity of 3 in the latter stems from the triplet states 
\begin{align}
\ket{\rm T_+}&=\ket{\uparrow\uparrow}\\
\ket{\rm T_0}&={1\over\sqrt{2}}\big(\ket{\uparrow\downarrow}+\ket{\downarrow\uparrow}\big)\\
\ket{\rm T_-}&=\ket{\downarrow\downarrow}
\end{align}
The coupled basis $\{\ket{\rm S},\ket{\rm T_+},\ket{\rm T_0},\ket{\rm T_-}\}$ describes the two unpaired electron spins. For example, for an RP containing $n$ spin-1/2 nuclei, it would be $d=2^{n+2}$  and the projectors would be written as $\qs=\ket{\rm S}\bra{\rm S}(\otimes\mathbbmtt{1}_{2})^n$ and $\qt=(\ket{\rm T_+}\bra{\rm T_+}+\ket{\rm T_0}\bra{\rm T_0}+\ket{\rm T_-}\bra{\rm T_-})(\otimes\mathbbmtt{1}_{2})^n$.

There are also two rates to consider, the singlet and triplet recombination rates, $\ks$ and $\kt$, respectively. These are defined as follows: at $t=0$ we prepare an RP ensemble having no magnetic interactions (${\cal H}=0$) in the singlet (triplet) electron state (and any nuclear spin state). Then its population would decay exponentially at the rate $\ks$ ($\kt$). In general, during a time interval $dt$, the measured singlet and triplet neutral products will be 
\begin{align}
dr_{\rm S}&=\ks dt\tr\{\rho\qs\}\label{drs}\\
dr_{\rm T}&=\kt dt\tr\{\rho\qt\}\label{drt}
\end{align}
Indeed, if all RPs were in the singlet or triplet state, the fraction of them recombining in the singlet or triplet channel during $dt$ would be $\ks dt$ and $\kt dt$, respectively. If they are in the general state described by $\rho$, then $\ks dt$ and $\kt dt$ have to be multiplied by the respective fraction of singlet and triplet RPs. 
\subsection{Initial state} 
Radical-ion pairs are usually produced from an electron spin zero neutral precursor, so their initial state is the electron singlet state. The thermal proton polarization at room temperature and at magnetic fields as large as 10 kG is about $10^{-5}$, which for all practical purposes can be approximated by zero. The RP initial state having zero nuclear polarization and being in the singlet electron spin state is  
\beq
\rho_0={\qs\over{\tr\{\qs\}}}\label{singlet}
\eeq
Indeed, since $\qs^2=\qs$, it is $\tr\{\rho_0\qs\}=1$. If $I_{i,j}$ is the $j$-th component ($j=x,y,z$) of the $i$-th nuclear spin, then $\tr\{\rho_0I_{i,j}\}=0$ since $I_{i,j}$ are traceless operators.
\begin{figure}
\begin{center}
\includegraphics[width=7 cm]{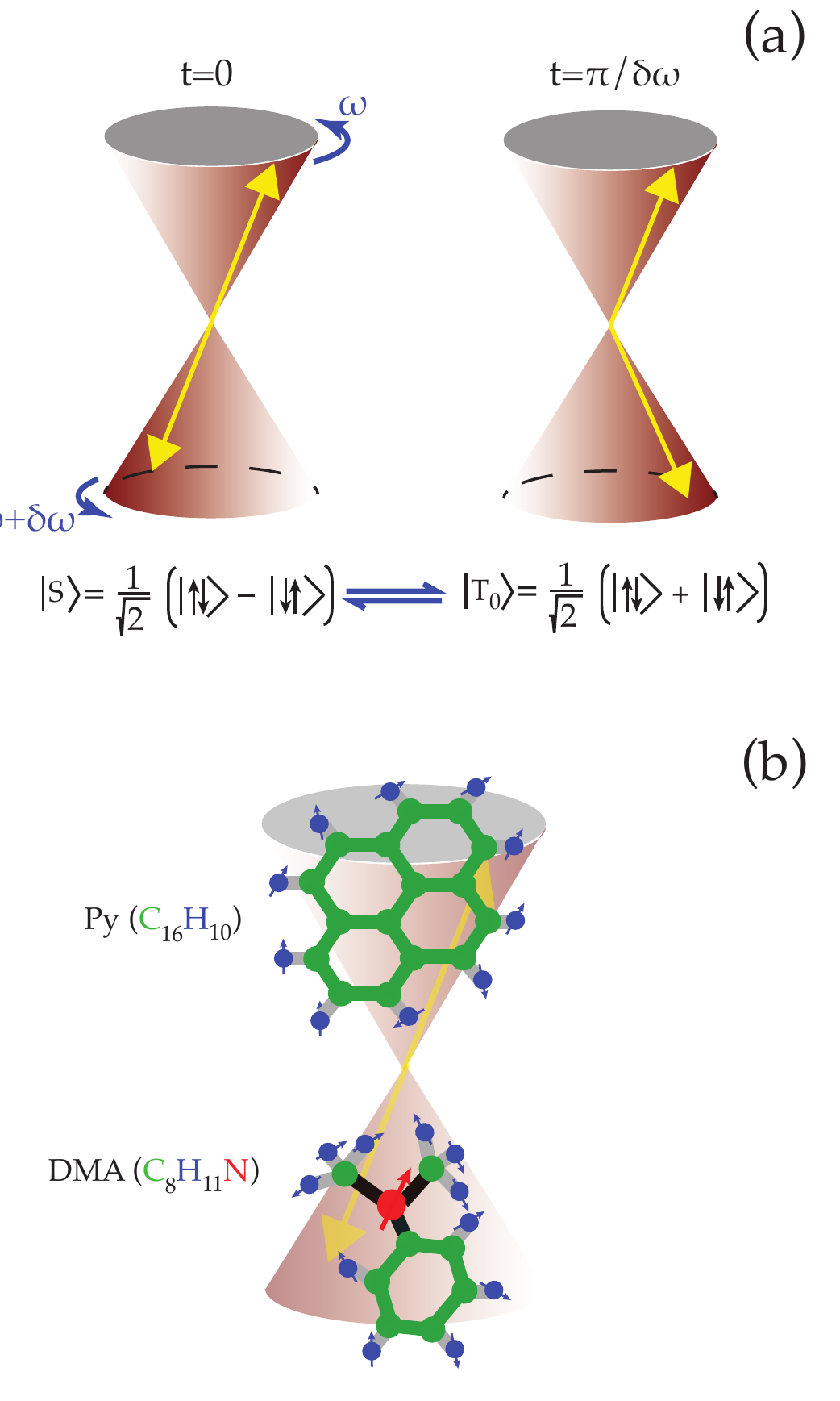}
\caption{(a) Classical vector model describing S-T conversion of a two-electron spin state driven by a difference in the electrons' Larmor frequencies. (b) Example of a radical-pair, ${\rm Py}^{\bullet +}{\rm DMA}^{\bullet -}$. The different nuclear spin environment of ${\rm Py}^{\bullet +}$ and ${\rm DMA}^{\bullet -}$ is the basis of singlet-triplet mixing.}
\label{ST2e}
\end{center}
\end{figure}
\subsection{Hamiltonian evolution}
The magnetic interactions included in ${\cal H}$ drive the unitary RP dynamics. The simplest way to understand S-T mixing is by using the classical vector model and considering first an imaginary RP consisting of just two electron spins having Larmor frequencies different by $\delta\omega$. As shown in Fig. \ref{ST2e}a, if the initial state is the singlet state $|{\rm S}\rangle$, after a time $\tau=\pi/\delta\omega$ the two spins will have developed a $\pi$ phase difference, their state becoming the triplet $|{\rm T}_0\rangle$. In reality this model is encountered in the so-called $\Delta g$ mechanism operating at high magnetic fields \cite{deltag}, where an actual difference in the g-factor of the two electrons is responsible for their different Larmor frequency. In most cases, however, S-T mixing is caused by the different nuclear spin environment coupled to each electron through hyperfine interactions. 
To illustrate hyperfine-induced S-T mixing, we consider the example of Py-DMA (Pyrene-Dimethylaniline), shown in Fig. \ref{ST2e}b. Here Py is the electron donor and DMA the acceptor, i.e. the RP is ${\rm Py}^{\bullet +}{\rm DMA}^{\bullet -}$. In this example, the donor radical has 10 proton spins hyperfine coupled to its unpaired electron, while the acceptor radical has 12 spins, 11 protons and one nitrogen, coupled to its unpaired electron. This difference in nuclear spin environments "seen" by the unpaired electrons of ${\rm Py}^{\bullet +}$ and ${\rm DMA}^{\bullet -}$ drives S-T mixing. 

To come back to the earlier discussion of the very possibility of spin-dependent chemical reactions, note that the diffusion constant of  Py/DMA in typical solvents used in \cite{schulten77} is $D=10^{-5}~{\rm cm^2/s}$, while the inter-radical distance where the RP recombination reaction is appreciable is \cite{efimova} on the order of 10 nm, hence the reaction dynamics take place at a timescale of about 100 ns. The electron spin precession time at a hyperfine magnetic field of 10 G is about 30 ns, short enough for spin motion to influence the reaction yields.

\begin{figure}
\begin{center}
\includegraphics[width=8.5 cm]{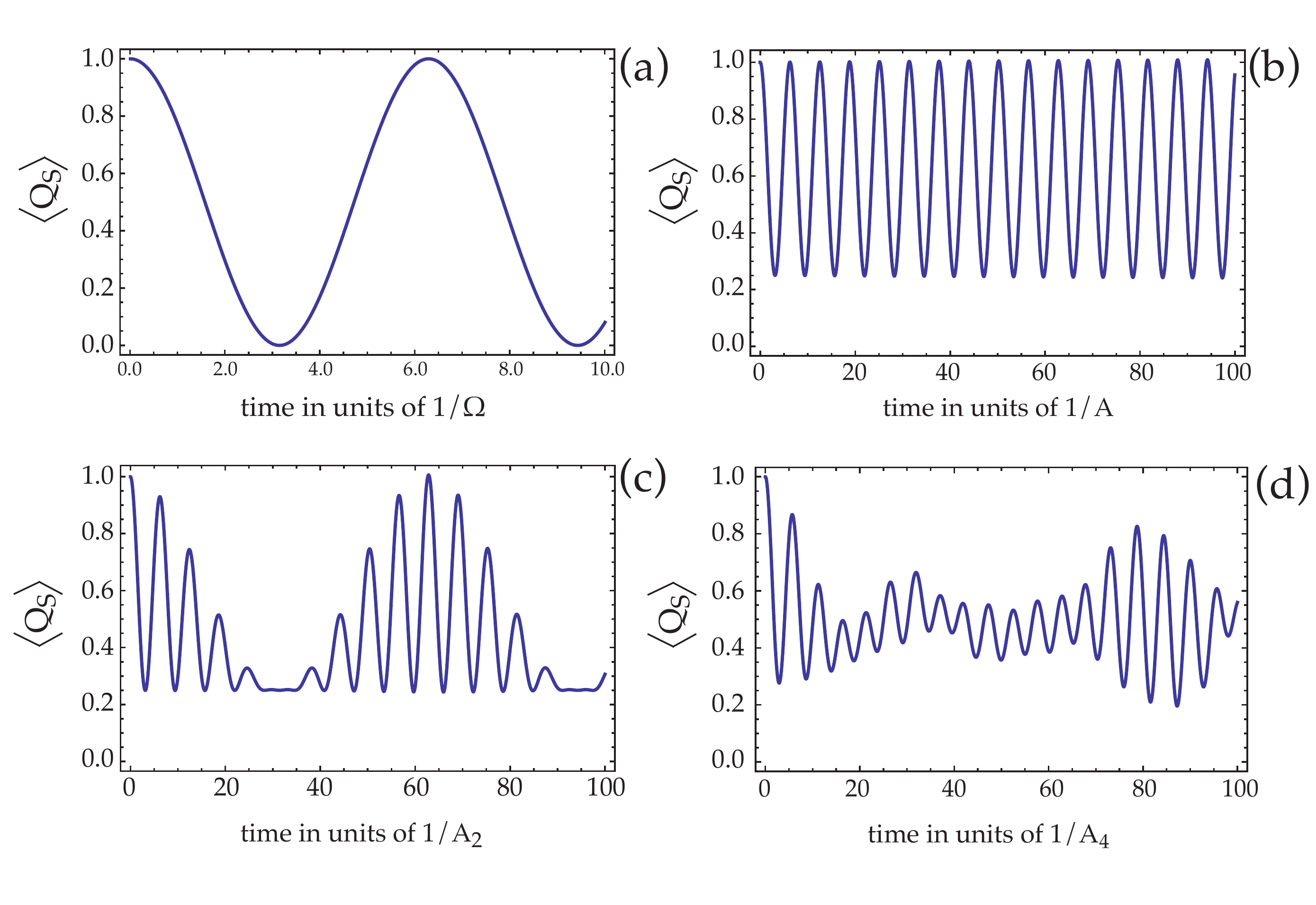}
\caption{Examples of S-T mixing driven by various magnetic Hamiltonians. (a) RP with just two electrons with different Larmor frequencies. (b) RP with one nuclear spin isotropically coupled to the donor's unpaired electron with hyperfine coupling $A$. (c) RP with one nuclear spin in each radical, isotropically coupled with hyperfine couplings $A_1$ and $A_2$. For the plot $A_2=10A_1$.  (d) RP with 4 nuclear spins in one radical and none in the other. The hyperfine couplings are $A_1=1$, $A_2=2$, $A_3=3$ and $A_4=10$.}
\label{examples}
\end{center}
\end{figure}

We will now visualize a few among many possibilities of S-T mixing, by plotting in Fig. \ref{examples} the Hamiltonian evolution of the RP spin state for four different Hamiltonians. 
We plot the expectation value of $\qs$ given by $\langle\qs\rangle=\tr\{\rho\qs\}$, where $\rho$ evolves just unitarily, $d\rho/dt=-i[{\cal H},\rho]$. In all cases we start with $\rho_0$ given in \eqref{singlet}. In Fig. \ref{examples}a we consider an RP with just two unpaired electron spins having different Larmor frequencies and residing in an external magnetic field along the z-axis, so that ${\cal H}=\omega_{D}s_{zD}+\omega_As_{zA}$. In this example $\rho$ is 4-dimensional. Singlet-triplet mixing occurs at the mixing frequency $\Omega=|\omega_D-\omega_A|$. In Fig. \ref{examples}b we consider an RP with just one spin-1/2 nucleus coupled to e.g. the donor electron, so the Hamiltonian reads ${\cal H}=A\mathbf{s}_D\cdot\mathbf{I}$, where $A$ is the isotropic hyperfine coupling and $\mathbf{I}$ the nuclear spin operator. In this case $d=8$. Moving to higher-dimensional examples, in Fig. \ref{examples}c we consider two spin-1/2 nuclei, one coupled to the donor and one to the acceptor electron. Now the Hamiltonian is ${\cal H}=A_1\mathbf{s}_D\cdot\mathbf{I}_1+A_2\mathbf{s}_A\cdot\mathbf{I}_2$, and $d=16$. Finally, in Fig. \ref{examples}d we consider a 64-dimensional case where 4 nuclear spins couple to the donor electron, hence ${\cal H}=\sum_{j=1}^{4}A_j\mathbf{s}_D\cdot\mathbf{I}_j$. Unlike the previous cases, the evolution is seen to be aperiodic \cite{tiersch}, due to a superposition of multiple S-T oscillations\footnote{In reality it is periodic, only the period is much larger than the timescale of the figure.}. Although not of concern for this review, we note that other sorts of interactions can be included in ${\cal H}$, like spin-exchange and long-range dipolar \cite{rodgers_review}. Moreover, hyperfine couplings can also be anisotropic \cite{horePRL}.
\subsection{Traditional master equation for radical-pair reactions}
Besides the unitary evolution of the RP spin density matrix $\rho$, the reaction super-operator describing the spin-dependent population loss into neutral products must also be taken into account. The majority of theoretical calculations were so far performed with Haberkorn's master equation, also termed "traditional" in recent years when other approaches came along:
\beq
{{d\rho}\over {dt}}=-i[{\cal H},\rho]-\underbracket{{\ks\over 2}\big(\qs\rho+\rho\qs\big)-{\kt\over 2}\big(\qt\rho+\rho\qt\big)}_{\rm Haberkorn's~reaction~super-operator}\label{trad}
\eeq
Obviously, the master equation for $\rho$ is fundamental for spin chemistry, being the basis for predicting all experimental observables. Surprisingly, a microscopic first-principles derivation of \eqref{trad} was given only recently \cite{ivanov} as a response to our work challenging it (see Section 6.2). As input, this or any contending master equation requires all magnetic interactions entering the Hamiltonian ${\cal H}$, and the two rates $\ks$ and $\kt$. As output, it can theoretically predict any experimental observable. A few examples often encountered are: (A) The singlet and triplet reaction yields follow by integrating \eqref{drs} and \eqref{drt}, $Y_{\rm x}=\int_{0}^{\infty}dr_{\rm x}$, with ${\rm x=S,T}$. Given that the initial density matrix is properly normalized, $\tr\{\rho_0\}=1$, it will be $Y_{\rm S}+Y_{\rm T}=1$. (B) The reaction rate, given by the decay rate of the RP population, $d\tr\{\rho\}/dt=-\ks\tr\{\rho\qs\}-\kt\tr\{\rho\qt\}$. In the special case $\ks=\kt\equiv k$, the RP population decays  exponentially at the rate $k$, i.e. $d\tr\{\rho\}/dt=-k\tr\{\rho\}$, which follows from the previous equation since\footnote{Henceforth the unit matrix $\mathbbmtt{1}$ has the dimension of the radical-pair under consideration.} $\qs+\qt=\mathbbmtt{1}$. (C) The magnetic field effect \cite{hore_nature} is about the magnetic field dependence of the time evolution of $\tr\{\rho\}$, and is defined as $MFE=\tr\{\rho\}_B-\tr\{\rho\}_{B=0}$. 

\begin{figure}
\begin{center}
\includegraphics[width=7.5 cm]{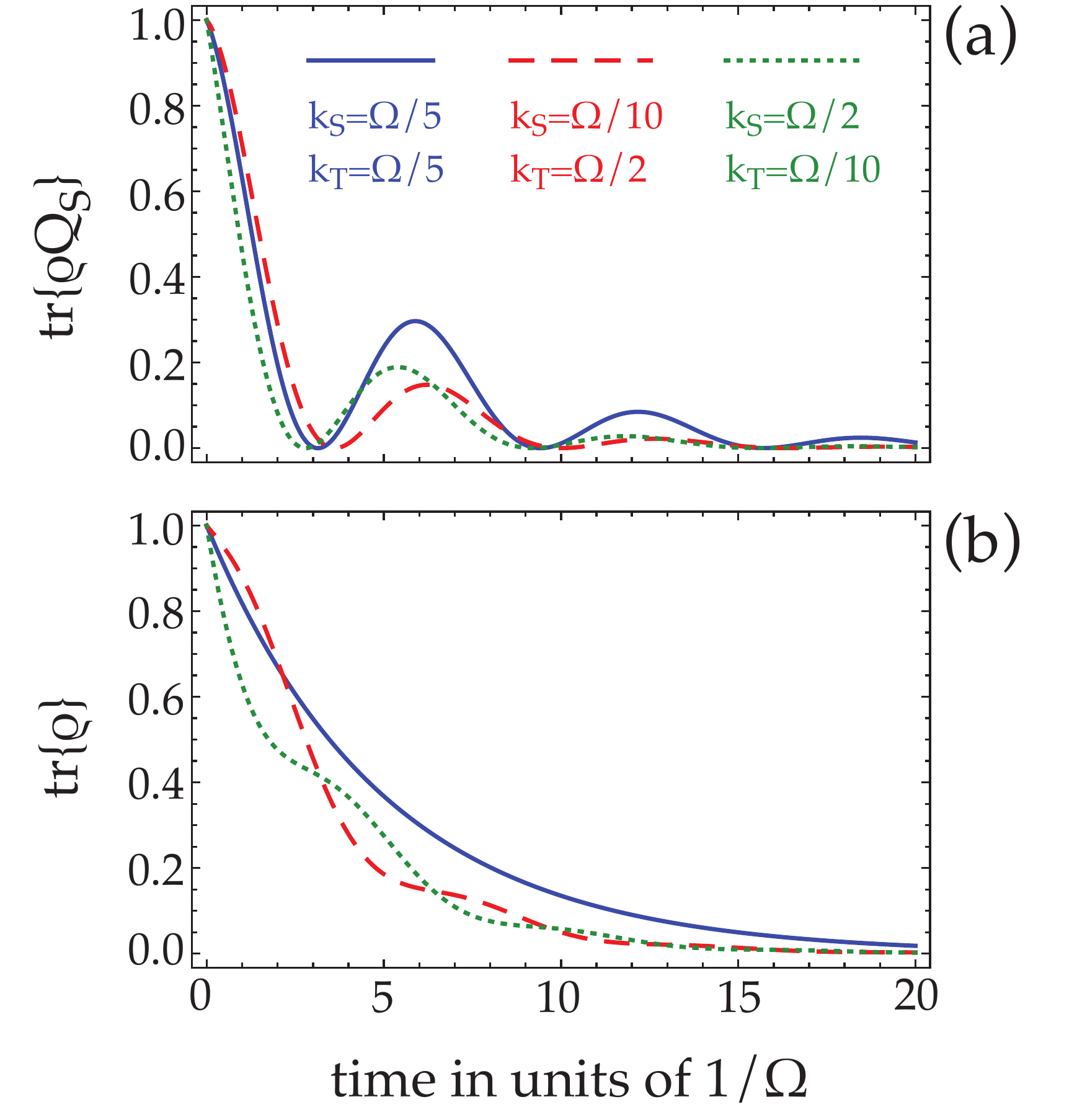}
\caption{For an imaginary RP with just two electrons and for the same Hamiltonian as in Fig.\ref{examples}a, we use the traditional master equation \eqref{trad} and  plot the time evolution of (a) $\langle\qs\rangle$ and (b) the RP population $\tr\{\rho\}$, for three different cases: (i) $\ks=\kt$, (ii) $\ks<\kt$ and (iii) $\ks>\kt$.}
\label{tradEV}
\end{center}
\end{figure}

To summarize, the particular form of the reaction super-operator of \eqref{trad} or any contending master equation is imprinted in all physically accessible observables.  Importantly, the reaction super-operator does not only reduce the RP population, $\tr\{\rho\}$, but also induces a change in the spin-state character embodied by $\rho$ when $\ks\neq\kt$. To understand this, consider a balanced mixture of singlet and triplet RPs which are magnetically frozen, i.e. there are no magnetic interactions, ${\cal H}=0$ (such idealized scenarios will be used frequently in order to gain physical intuition on the dynamics). If $\ks=\kt$, the balance is not distorted as the RPs disappear at the same rate irrespective of their total electron spin. On the other hand, if $\ks\neq\kt$, the ensemble of surviving RPs will become more singlet (triplet) if $\kt>\ks$ ($\kt<\ks$). As an example, we plot in Fig. \ref{tradEV} the time dependence of (a) the singlet character of $\rho$, $\tr\{\rho\qs\}$, and (b) the RP population, $\tr\{\rho\}$, for the Hamiltonian used in Fig. \ref{examples}a, but now calculated from the full master equation \eqref{trad}. It is clearly seen that the time evolution of both observables depends on the relative magnitude of $\ks$ and $\kt$, and of course on the particular form of the reaction super-operator in \eqref{trad}.
\section{Quantum foundations of the radical-pair mechanism}
We will now lay out the main premise of this review, demonstrating that the RPM is a biochemical system surprisingly rich in quantum-information-science concepts and effects. The need for such concepts appears when one recognizes the physical model behind the RPM and attempts to translate this model into a master equation. 
\subsection{Physical model of the radical-pair mechanism}
\begin{figure}
\begin{center}
\includegraphics[width=8.5 cm]{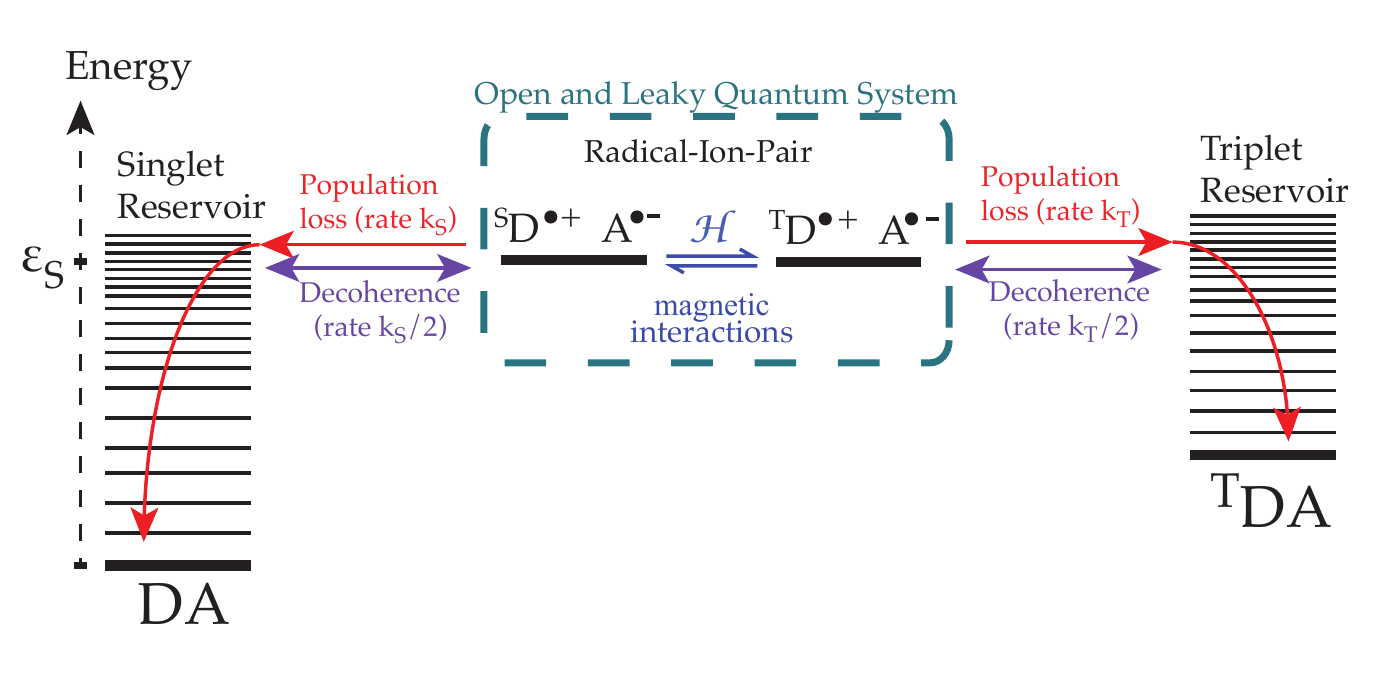}
\caption{Radical-pair energy levels including the vibrational excitations of the neutral product states. Due to these reservoir states, RPs (i) lose population by recombination into the neutral ground states through {\it two real} and consecutive transitions, from the RP to the quasi-resonant reservoir states followed by the decay of the latter to the ground state, and (ii) simultaneously suffer S-T decoherence due to virtual transitions from the RP to the reservoir states {\it and back}.}
\label{rip_levels}
\end{center}
\end{figure}

\begin{figure}
\begin{center}
\includegraphics[width=8 cm]{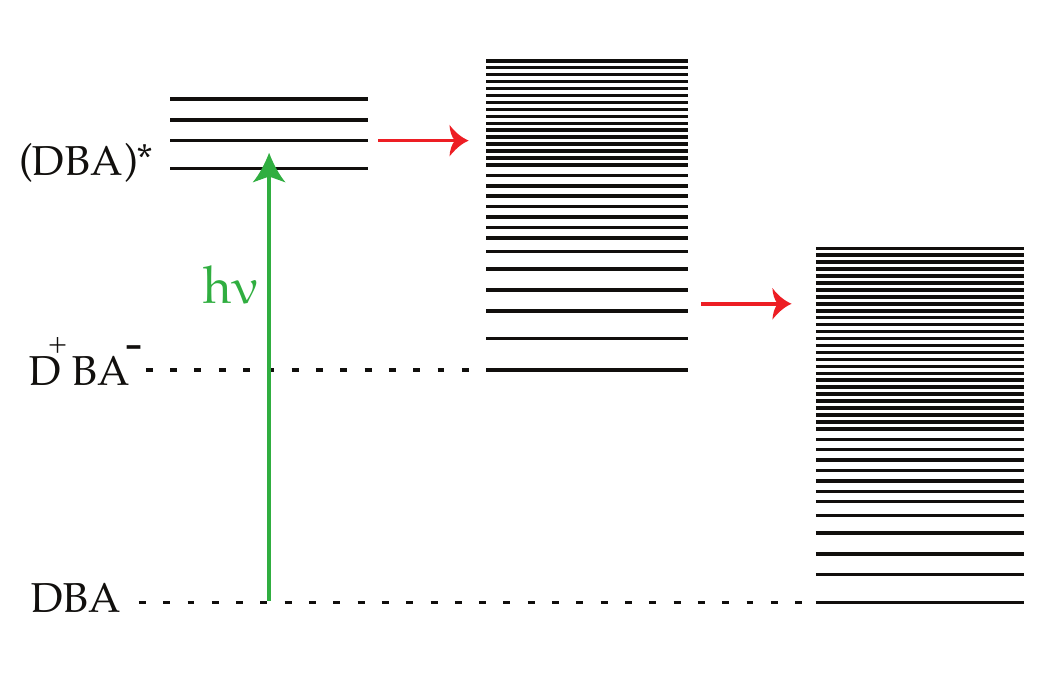}
\caption{Reproduction of Fig. 1 of \cite{bixon_jortner}, where $h\nu$ is the photon energy exciting the donor-bridge-acceptor DBA to (DBA)$^*$, and the horizontal red arrows denote the intramolecular charge transfers, the creation of the charge separated state ${\rm D^{+}BA^{-}}$ (1), and the recombined state DBA (2).}
\label{jortner}
\end{center}
\end{figure}

The diagrams depicting the reaction dynamics (Fig. \ref{schematic}a and Fig. \ref{schematic}b) convey a gross picture but miss important details. In Fig. \ref{rip_levels} we present a more detailed energy level structure that is critical in unraveling the quantum foundations of RPM.
We brought into the picture the vibrational excitations of the neutral product molecules. At high excitation energy, these states form a quasi-continuous reservoir quasi-resonant with the RP energy levels. The physics resulting from this reservoir are at the core of our work. Before proceeding, we should note that the level structure presented in Fig. \ref{rip_levels} is typical for electron-transfer studies \cite{et1,et2,et3,et4} and was well known before our work. For example, in Fig. \ref{jortner} we reproduce a figure appearing in a study on supermolecular charge separation \cite{bixon_jortner}. The authors state (Section II of \cite{bixon_jortner}) that {\it charge recombination ${\rm D^{+}BA^{-}\rightarrow DBA}$ is realized with a vibronic level of the charge-transfer state decaying into a vibrationally excited ground-state manifold}. What we have done in Fig. \ref{rip_levels} is to adapt exactly this model and also include spin conservation by considering two different manifolds for the general case of singlet and triplet recombination products. 

While the model of Fig. \ref{rip_levels} was understood before our work, the physical picture following from taking this model at face value was not. 
We have shown that radical-pairs are both {\it leaky} and {\it open} quantum systems, hence their non-unitary dynamics consist of two processes unfolding simultaneously and independently: (a) singlet-triplet decoherence and (b) spin-dependent population decay. Both processes are intimately linked to the vibrational reservoirs.
\subsection{Quantum phononics: the role of phonon vacuum in radical-pair reactions}
The phonon degrees of freedom in RP reactions are reminiscent of another emerging field, quantum phononics. Coherent manipulation of phonons \cite{mahboob} is witnessing a rapid growth in recent years both due to the fundamental interest of reaching the quantum limit in mechanical degrees of freedom \cite{vahala} and due to the potential of developing novel phonon-based technologies \cite{rmp}. Regarding biological systems, it would at first sight appear that vibrations just represent a classical noise source adding to the complexity of the molecular scaffold supporting biological dynamics. Again, the counterintuitive picture is currently emerging that nature is also able to harness quantum degrees of freedom of vibrations in intricate ways. In photosynthesis, phonons are understood to crucially regulate energy transport in light-harvesting complexes \cite{guzik,plenioDAT_jcp,plenioDAT,castro1,castro2}. What we will demonstrate next is that the {\it phonon vacuum} is the fundamental source of S-T decoherence in RP reactions, much like the photon vacuum is for atomic coherences.
\subsection{Singlet-triplet decoherence in the radical-pair mechanism: null quantum measurement of recombination products}
Null quantum measurements are well-known in quantum optics. A vivid example \cite{haroche} is the single atom surrounded by an array of unit-quantum-efficiency photodetectors. If the atom is initially in a coherent superposition of the ground and excited states, $|\psi_0\rangle=(|g\rangle+|e\rangle)/\sqrt{2}$, and if no photon is detected until time $t$, the atomic state will be (apart from a normalization constant) $|\psi_t\rangle\propto e^{-\Gamma t/2}|e\rangle+|g\rangle$, where $\Gamma$ is the spontaneous decay rate. The interpretation is that even the absence of a photon detection can condition our knowledge of the atom's state, e.g. if no photon is ever detected then the atom's state cannot contain an excited state component.

In analogy, consider a single radical-ion pair in the spin state $|\psi_0\rangle$ at $t=0$. Suppose that within the following time interval $dt$ there is no detection of a neutral recombination product. What is $|\psi_{dt}\rangle$ ? In our first publication \cite{pre2009} we provided a first-principles derivation of the master equation for non-recombining (or surviving) RPs, that is, the case of no detection of recombination products.

We have shown that until an RP charge-recombines, it does not just undergo Hamiltonian evolution, as was intuitively expected, but it suffers random projections to the singlet or triplet electron spin state. In the ensemble picture, this represents a {\it fundamental} S-T decoherence process in the sense that it cannot in any way be mitigated. In contrast, {\it technical} decoherence can in principle be suppressed in an experiment designed carefully enough. To give an exaggerated example, inadvertently radiating RPs with radio-frequency noise (as it is purposefully done in studying bird disorientation \cite{mouritsen}), would cause spin decoherence. This can obviously be suppressed by removing rf noise sources from the experiment's proximity, hence rf noise is an example of technical decoherence. The physical process we will present next is not such a technical source of decoherence, but an unavoidable one, intimately linked to the radical-pair recombination mechanism.
\subsection{Derivation of the master equation describing S-T decoherence}
When dealing with open quantum systems it is crucial to correctly identify the open system and its environment, so that one can apply the standard treatment of evolving the combined system and tracing out the environment. Based on Fig. \ref{rip_levels}, we have identified as the "system" the spin degrees of freedom of the radical-ion pair, described by the RP density matrix $\rho$. The "environment" is represented by the two reservoirs, the excited vibrational manifolds of the neutral ground states DA and $^{\rm T}$DA. We stress that these vibrational states are empty, or unoccupied, as long as the RP has not recombined. They should not be confused with the thermally excited vibrational states of the radical-pair itself. When the back electron-transfer from the acceptor A to the donor D neutralizes the radical-pair, the neutral molecule finds itself in an excited vibrational state. These are the reservoir states, and their quantum state during the RP's lifetime is the vacuum. 

The state evolution of surviving RPs was based on a $2^{\rm nd}$-order perturbation treatment of the spin-selective RP coupling with the {\it unoccupied} vibrational reservoirs, presented in \cite{pre2009} and from a slightly different perspective in \cite{lamb}, the main points of which we recapitulate here. Consider for the moment just the singlet reservoir. Since the amplitude for the transition to the $i$-th singlet reservoir state, ${\rm D}^{\bullet +}{\rm A}^{\bullet -}\rightarrow{\rm ^{\rm S}DA}_{i}^{*}$, should be proportional to the singlet character of the RP state, the coupling Hamiltonian reads ${\cal V}=\sum_{i}\big(h_{i}+h_{i}^{\dagger}\big)$, where $h_{i}={u_{i}a_{i}^{\dagger}c\qs}$. The operator $\qs$ projects the RP state onto the electron-singlet subspace, while the raising operator $a_{i}^{\dagger}$ produces a single occupation of the $i$-th reservoir level. The transition amplitude is $u_i$, and the operator $c$ describes the occupation of the acceptor's electron site, i.e. $c^{\dagger}c=1$ ($c^\dagger c=0$) means the electron is localized at the acceptor (donor). The hermitian conjugate $h_{i}^{\dagger}$ describes the reverse process ${\rm ^{\rm S}DA}^{*}_{i}\rightarrow$ $^{\rm S}{\rm D}^{\bullet +}{\rm A}^{\bullet -}$. As explained in \cite{pre2009,pre2011}, virtual transitions ${\rm D}^{\bullet +}{\rm A}^{\bullet -}\rightarrow{\rm ^{\rm S}DA}^{*}_{i}\rightarrow$ $^{\rm S}{\rm D}^{\bullet +}{\rm A}^{\bullet -}$ driven by $h_{i}$ followed by $h_{i}^{\dagger}$, produce the random singlet projections. 
Tracing out the reservoir degrees of freedom we arrive at (tilde denotes interaction-picture):
\begin{align}
{d\tilde{\rho}(t)\over{dt}}=&\sum_{ij}\int_{0}^{\infty}d\tau~\tilde{e}_{ij} e^{-i\epsilon_{S}\tau}
\nonumber\\&\Big[\tilde{\rm Q}_{S}(t)\tilde{\rho}(t)\tilde{\rm Q}_{S}^{\dagger}(t-\tau)-\tilde{\rho}(t)\tilde{\rm Q}_{S}^{\dagger}(t-\tau)\tilde{\rm Q}_{S}(t)\Big]\nonumber\\&+{\rm h.c.}\label{tilder},
\end{align}
where we also traced out the $c$ degrees of freedom as we are interested in the time evolution of the RP state, for which $\langle c^{\dagger}c\rangle=1$. The reservoir operators are $\tilde{E}_{i}=u_{i}e^{i\epsilon_{i}t}a_{i}^{\dagger}$, and the expectation value $\tilde{e}_{ij}=\langle E_{i}(0)\tilde{E}_{j}^{\dagger}(\tau)\rangle$ is calculated in the phonon vacuum state. Finally, the transition amplitudes $u_j$ are composed of an electronic matrix element and a nuclear overlap matrix element, $u_j=v_{j}\chi_{j}$ \cite{jortnerET}. We consider these to be independent of $j$ in the vicinity of $\epsilon_{S}$ (see Fig. 5), and introduce the Franck-Condon averaged density of states $g(\epsilon)=|\chi(\epsilon)|^2d(\epsilon)$, which takes into account both $\chi(\epsilon)$, the nuclear wave function overlap integral, and $d(\epsilon)$, the density of vibrational states at the energy $\epsilon$. Identifying the recombination rate $\ks$ with the expression $2\pi|v|^{2}g(\epsilon_{S})$ we obtain $d\rho/dt=-{\ks\over 2}(\qs\rho+\rho\qs-2\qs\rho\qs)$. Adding into the picture the triplet reservoir and the unperturbed unitary evolution, we arrive at the Lindblad-type master equation
\begin{align}
{{d\rho}\over {dt}}\Big |_{\rm decoh}=&-i[{\cal H},\rho]\nonumber\\&-{\ks\over 2}(\qs\rho+\rho\qs-2\qs\rho\qs)\nonumber\\&-{\kt\over 2}(\qt\rho+\rho\qt-2\qt\rho\qt)
\end{align}
Due to the relation $\qs+\qt=\mathbbmtt{1}$, the above can take the more compact form 
\beq
{{d\rho}\over {dt}}\Big |_{\rm decoh}=-i[{\cal H},\rho]+{\cal D}\llbracket\rho\rrbracket,\label{MEnr}
\eeq
where the S-T decoherence super-operator is
\beq
{\cal D}\llbracket\rho\rrbracket=-{{\ks+\kt}\over 2}(\qs\rho+\rho\qs-2\qs\rho\qs)\label{Drho}
\eeq
\subsection{Null quantum measurement of recombination products in terms of single-molecule quantum trajectories}
Typically, a Lindblad-type equation describes the effect of a continuous quantum measurement \cite{jacobs} on the system's density matrix. What is measured in this case and who performs the measurement? The measured observable is the singlet character of the RP's spin state, i.e. the observable $\qs$. The measurement is performed by the molecule's own vibrational reservoir, i.e. by the existence of the phonon vacuum and the possibility the RP has to occupy these phonon reservoir states. This is an intra-molecule continuous quantum measurement. The density matrix evolving according to \eqref{MEnr} describes an ensemble of radical-pairs. What about the equivalent picture in terms of quantum trajectories? Consider an RP in the pure state $|\psi\rangle$ at time $t$. Assume that no recombination product was observed within $dt$. What is $|\psi\rangle$ at time $t+dt$? There are three different possibilities for single-molecule state evolution:\newline
{\bf (i)} projection to the singlet RP state 
\begin{equation*}
|\psi_{\rm S}\rangle={{\qs|\psi\rangle}\over \sqrt{\langle\psi|\qs|\psi\rangle}},
\end{equation*}
taking place with probability $dp_{\rm S}={{(k_{\rm S}+k_{\rm T})dt}\over 2}\langle\psi|\qs|\psi\rangle$.
{\bf (ii)} projection to the triplet RP state 
\begin{equation*}
|\psi_{\rm T}\rangle={{\qt|\psi\rangle}\over \sqrt{\langle\psi|\qt|\psi\rangle}},
\end{equation*}
taking place with probability $dp_{\rm T}={{(k_{\rm S}+k_{\rm T})dt}\over 2}\langle\psi|\qt|\psi\rangle$.
{\bf (iii)} unitary evolution driven by ${\cal H}$,  taking place with probability $1-dp_{\rm S}-dp_{\rm T}$.

In an ensemble of RPs, these single-molecule events are unobservable, so we have to average over (i)-(iii), transforming an initially pure into a mixed state.
This average {\it exactly} reproduces the master equation \eqref{MEnr}. In other words, writing 
\begin{align}
\rho_{t+dt}&=dp_{\rm S}\rho_{\rm S}\nonumber\\&+dp_{\rm T}\rho_{\rm T}\nonumber\\&+(1-dp_{\rm S}-dp_{\rm T})(\rho_t-idt[{\cal H},\rho_t]),
\end{align}
leads to \eqref{MEnr} for $d\rho/dt={{\rho_{t+dt}-\rho_t}\over{dt}}$, $\rho_t=|\psi\rangle\langle\psi |$,  $\rho_{\rm S}=|\psi_{\rm S}\rangle\langle\psi_{\rm S}|$ and $\rho_{\rm T}=|\psi_{\rm T}\rangle\langle\psi_{\rm T}|$. 
 
The physical significance of the sum $k_{\rm S}+k_{\rm T}$ appearing in the probabilities $dp_{\rm S}$ and $dp_{\rm T}$ and also appearing in \eqref{MEnr} is the fact that both singlet and triplet reservoirs continuously "measure" the same observable, namely ${\rm Q_S}$. The result of this measurement is either 1 or 0, corresponding to the singlet and triplet projections, respectively. In particular, the singlet reservoir measures the observable ${\rm Q_S}$ at the rate $k_{\rm S}/2$. The "yes" result of this measurement corresponds to $\qs\mapsto 1$ and the singlet projection, while the no/null result corresponds to $\qs\mapsto 0$ and the triplet projection. Similarly, the triplet reservoir measures the observable ${\rm Q_T}=\mathbbmtt{1}-{\rm Q_S}$ at the rate $k_{\rm T}/2$. The "yes" result of this measurement corresponds to $\qs\mapsto 0$ and a triplet projection, while the no/null result corresponds to $\qs\mapsto 1$ and the singlet projection. Equivalently, ${\rm Q_S}$ is measured at the total rate $(k_{\rm S}+k_{\rm T})/2$. These measurement results are unobservable, each radical-pair has its own random history of projections, hence the ensemble exhibits a decay of S-T coherence. 
\begin{figure}
\begin{center}
\includegraphics[width=8 cm]{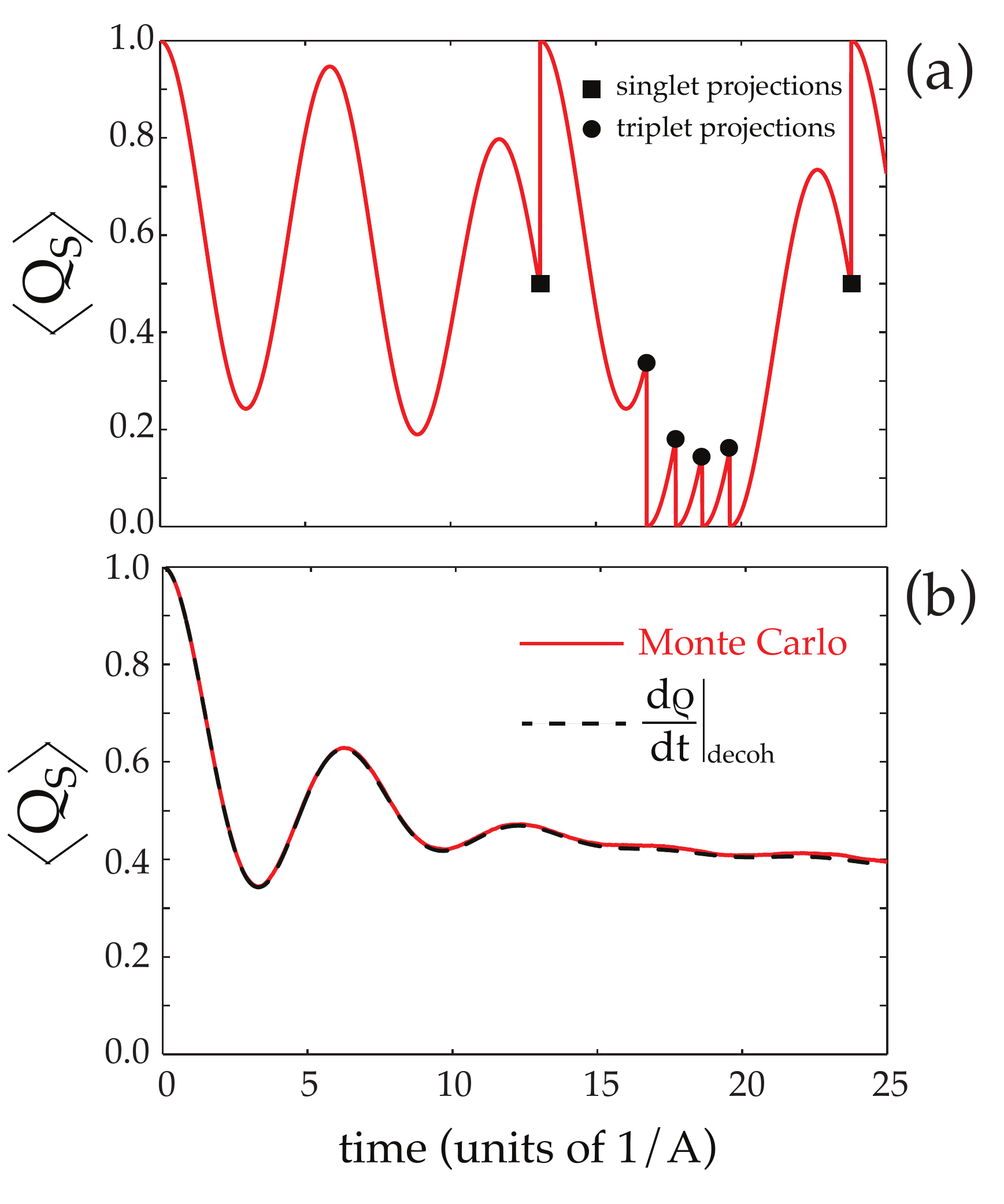}
\caption{Time evolution of $\langle{\rm Q_S}\rangle$ for a model RP with one nuclear spin, taking into account only S-T decoherence and S-T mixing driven by ${\cal H}=\omega(s_{Dz}+s_{Az})+A\mathbf{s}_{D}\cdot\mathbf{I}$. We took $\omega=A/10$ and $\ks=\kt=A/4$. (a) single-RP quantum trajectory, depicting singlet and triplet projections at random instants in time. The initial RP state for this trajectory is $\ket{\rm S}\otimes\ket{\Uparrow}$. (b) average of 20,000 such trajectories (red solid line), half of which have initial state $\ket{\rm S}\otimes\ket{\Uparrow}$ while the other half have initial state $\ket{\rm S}\otimes\ket{\Downarrow}$. The prediction of the master equation \eqref{MEnr} is shown by the black dashed line. The initial state of the density matrix was $\rho_0$ given by \eqref{singlet}.}
\label{decoh}
\end{center}
\end{figure}
In Fig. \ref{decoh}b we plot an example (reproduced from \cite{pre2014}) of the evolution described by \eqref{MEnr}. The decaying amplitude of the $\langle\qs\rangle$-oscillations is what this decoherence describes. A single-RP quantum trajectory is shown in Fig.\ref{decoh}a. This trajectory consists of the Hamiltonian evolution interrupted at random instants by the aforementioned singlet or triplet projections. Averaging a large number of such trajectories exactly reproduces the evolution of Fig. \ref{decoh}b.
\subsection{The two roles of the coupling Hamiltonian}
In contrast to S-T dephasing produced by {\it virtual} transitions to the reservoir states and back, charge recombination proceeds by {\it real} transitions to the reservoir states, followed by their decay to the ground state. The latter happens at ps timescales, so the rate limiting process is the former. Within 1$^{\rm st}$-order perturbation theory and by using Fermi's golden rule we find that the recombination rate will be $k_{S}=2\pi |\langle f|{\cal V}|i\rangle|^{2}d(\epsilon_{S})=2\pi|v|^2g(\epsilon_{S})$, where as initial state $|i\rangle$ we chose a pure singlet state of the radical-pair, and as a final state $|f\rangle$ one among the near-resonant and quasi-continuum reservoir states described by the density of states $d(E)$. So the recombination rates calculated from Fermi's golden rule have the exact same form {\it identified} in the previous derivation of ${{d\rho}\over {dt}}\big |_{\rm decoh}$. This is not a coincidence, since the Hamiltonian ${\cal V}$ coupling the radical-pair with the vibrational reservoir is responsible for both processes, which are of order $|{\cal V}|^2$. Indeed, the coupling ${\cal V}$ appears two times in \eqref{tilder}, and the rate $k_{\rm S}$ depends on the square of ${\cal V}$ through Fermi's golden rule. Hence both processes lead to first-order reaction kinetics, $d\rho\propto dt$. 
 \subsection{Reaction terms}
The master equation \eqref{MEnr} is trace-preserving, i.e. under the evolution ${\cal D}\llbracket\rho\rrbracket$ the normalization $\tr\{\rho\}$ remains constant. We will now describe the second process, the spin-dependent recombination, leading to neutral reaction products and the concomitant decay of $\tr\{\rho\}$. However, charge recombination does not only lead to a decay of $\tr\{\rho\}$. Radical-ion pairs are in some quantum state at the moment of their recombination, hence their disappearance from the RP ensemble leads to yet another state change of the ensemble of surviving RPs (see Section 2.5). 

Given $\ks$ and $\kt$, we can calculate the number of RPs that recombine during $dt$ in the singlet and triplet channel. These are $dr_{\rm S}=\ks dt\tr\{\rho\qs\}$ and $dr_{\rm T}=\kt dt\tr\{\rho\qt\}$, already introduced in \eqref{drs} and \eqref{drt}. Equivalently, $dr_{\rm S}$ ($dr_{\rm T}$) is the probability that a single RP will recombine during $dt$ in the singlet (triplet) channel.
However, just knowing these probabilities does not tell us what the state of the RP was before it recombined. Put differently, the {\it physical information we have at hand} is the detection of neutral recombination products (the ground states DA or $^{\rm T}$DA of Fig. 5), the number of which reflect these probabilities. Based on this information, we have to find the pre-recombination state of the RPs recombined during $dt$.

With this discussion we motivate the formal calculation \cite{pre2014} of the reaction terms. This is accomplished using a number of quantum-information concepts. First, we define a quantum coherence quantifier, a measure mapping the RP density matrix $\rho$ into a number quantifying S-T coherence. Second, we need the concept of quantum retrodiction, developed in the theory of quantum communications. 
\subsubsection{Radical-pair recombination, S-T coherence, and quantum retrodiction}
Quantum retrodiction \cite{retro1,retro2,retro3,retro4,molmer} is the reverse of prediction, which we regularly deal with. That is, given a density matrix describing a physical system, it is straightforward to predict the probabilities of possible measurement outcomes. What is less straightforward is to {\it retrodict} the system's preparation state given a specific measurement outcome. This is relevant to quantum communications \cite{jones_book}, since Bob, the receiving end of a communicating party, attempts to reconstruct the quantum state delivered to him by Alice, the sender, based on specific outcomes of quantum measurements he decides to perform. 

To demonstrate the need of quantum retrodiction in RP reactions, consider an RP ensemble and a time interval $dt$ so small that only one RP recombines during $dt$. If we knew the state $|\psi\rangle$ of this RP just before it recombines, we would subtract $|\psi\rangle\langle\psi|$ from the ensemble density matrix to account for this recombination event. But at any given moment, the density matrix of an RP ensemble consists of a mixture of pure states that have been produced by the magnetic Hamiltonian ${\cal H}$, but also undergone a number of singlet or triplet projections at random and unknown times. Given the neutral recombination product, which is either the singlet or the triplet ground state, one cannot unambiguously retrodict the pre-recombination state $|\psi\rangle$. The theory of quantum retrodiction allows us to probabilistically do so, in a way that depends on S-T coherence of the RP ensemble at the time we detect the particular recombination products. 

To give a few simple examples of why this is so, consider a maximally incoherent mixture of $n$ singlet and $m$ triplet RPs, described by $\rho_t=n|{\rm S}\rangle\langle{\rm S}|+m|{\rm T}_0\rangle\langle{\rm T}_0|$, the normalization being $\tr\{\rho_t\}=n+m$. {\it With this information at hand}, and given the detection of e.g. a singlet neutral product within $dt$, it is certain that the neutral product originated from an RP in the state $|{\rm S}\rangle\langle{\rm S}|$. To account for this recombination event, one would simply use the reaction term $d\rho=-|{\rm S}\rangle\langle{\rm S}|$, and hence $\rho_{t+dt}=\rho_{t}+d\rho=(n-1)|{\rm S}\rangle\langle{\rm S}|+m|{\rm T}_0\rangle\langle{\rm T}_0|$. This mixture had zero S-T coherence. Another example is a mixture of $n$ radical-pairs in the state $|\psi_1\rangle=(|{\rm S}\rangle+|{\rm T}_0\rangle)/\sqrt{2}$ and $m$ radical-pairs in the singlet state $|\psi_2\rangle=|{\rm S}\rangle$. This mixture has partial S-T coherence.  Assume that within $dt$ we detect a singlet neutral product. Did this originate from an RP in the state $|\psi_1\rangle$ or $|\psi_2\rangle$? Intuitively one could argue that it is the former with probability $n/(n+2m)$ and the latter with probability $2m/(n+2m)$. But how do we find the answer when we are just given a density matrix $\rho$?  
Given an arbitrary density matrix $\rho$, we can left and right multiply $\rho$ with $\mathbbmtt{1}=\qs+\qt$ and thus write 
\beq
\rho=\rho_{\rm SS}+\rho_{\rm TT}+\rho_{\rm ST}+\rho_{\rm TS}\label{decomp},
\eeq
where $\rho_{xy}=Q_{x}\rho Q_{y}$, with $x={\rm S,T}$. The last two terms, $\rho_{\rm ST}+\rho_{\rm TS}$, embody S-T coherence. In fact, the decoherence super-operator ${\cal D}\llbracket\rho\rrbracket$ exactly damps $\rho_{\rm ST}+\rho_{\rm TS}$, i.e. $\qs\rho+\rho\qs-2\qs\rho\qs=\rho_{\rm ST}+\rho_{\rm TS}$. In other words, virtual transitions ${\rm RP}\rightarrow{\rm vibrational~reservoir}\rightarrow{\rm RP}$ push $\rho$ towards an incoherent mixture $\rho_{\rm SS}+\rho_{\rm TT}$. The extent to which this is accomplished also depends on how fast S-T coherence is generated by the Hamiltonian ${\cal H}$.

It is the closure relation $\mathbbmtt{1}=\qs+\qt$ that automatically led to \eqref{decomp}, forced upon the RP dynamics by angular momentum conservation, which breaks up the RP's state space into eigenstates of either $\qs$ or $\qt$. On the other hand, the magnetic Hamiltonian ${\cal H}$ mixes the eigenstates of $\qs$ and $\qt$, producing S-T coherence. Which are the density matrix elements describing S-T coherence in the term $\rho_{\rm ST}+\rho_{\rm TS}$? To address this question, we note that $\rho$ also describes the spin state of a (possibly large) number of nuclear spins. Hence $\rho_{\rm ST}+\rho_{\rm TS}$ is in general a block-off-diagonal matrix. We will give a few simple examples to elucidate this point. Considering first an imaginary RP with just two electrons, the density matrix $\rho$ (of dimension 4) and it's coherent part, $\rho_{\rm ST}+\rho_{\rm TS}$, written in the basis $\{\ket{\rm S},\ket{\rm T_+},\ket{\rm T_0},\ket{\rm T_-}\}$, would be 
\begin{align}
\rho&=\begin{pmatrix}
\rho_{\rm SS} & \rho_{\rm ST_+} &  \rho_{\rm ST_0} &  \rho_{\rm ST_-}\\
\rho_{\rm T_+S} & \rho_{\rm T_+T_+} &  \rho_{\rm T_+T_0} &  \rho_{\rm T_+T_-}\\
\rho_{\rm T_0S} & \rho_{\rm T_0T_+} &  \rho_{\rm T_0T_0} &  \rho_{\rm T_0T_-}\\
\rho_{\rm T_-S} & \rho_{\rm T_-T_+} &  \rho_{\rm T_-T_0} &  \rho_{\rm T_-T_-}\\
\end{pmatrix}\nonumber\\\rho_{\rm ST}+\rho_{\rm TS}&=\begin{pmatrix}
0 & \rho_{\rm ST_+} &  \rho_{\rm ST_0} &  \rho_{\rm ST_-}\\
\rho_{\rm T_+S} & 0&  0& 0\\
\rho_{\rm T_0S} & 0 & 0 & 0\\
\rho_{\rm T_-S} &0 & 0& 0\\
\end{pmatrix}\label{rho4}
\end{align}
For an 8-dimensional radical-pair with one spin-1/2 nucleus the basis is \{$\ket{\rm S}\otimes\ket{\Uparrow}$,$\ket{\rm S}\otimes\ket{\Downarrow}$,$\ket{\rm T_+}\otimes\ket{\Uparrow}$,$\ket{\rm T_+}\otimes\ket{\Downarrow}$,$\ket{\rm T_0}\otimes\ket{\Uparrow}$,$\ket{\rm T_0}\otimes\ket{\Downarrow}$,$\ket{\rm T_-}\otimes\ket{\Uparrow}$,$\ket{\rm T_-}\otimes\ket{\Downarrow}$\}, and
\begin{widetext}
\beq
\rho_{\rm ST}+\rho_{\rm TS}=\begin{pmatrix}
0 & 0 & \rho_{\rm ST_+\Uparrow\Uparrow} &  \rho_{\rm ST_+\Uparrow\Downarrow} &  \rho_{\rm ST_0\Uparrow\Uparrow}& \rho_{\rm ST_0\Uparrow\Downarrow} &  \rho_{\rm ST_-\Uparrow\Uparrow} &  \rho_{\rm ST_-\Uparrow\Downarrow} \\
0 & 0 & \rho_{\rm ST_+\Downarrow\Uparrow} &  \rho_{\rm ST_+\Downarrow\Downarrow} &  \rho_{\rm ST_0\Downarrow\Uparrow}& \rho_{\rm ST_0\Downarrow\Downarrow} &  \rho_{\rm ST_-\Downarrow\Uparrow} &  \rho_{\rm ST_-\Downarrow\Downarrow}\\
\rho_{\rm T_+S\Uparrow\Uparrow} & \rho_{\rm T_+S\Uparrow\Downarrow}&  0& 0 & 0 & 0 & 0& 0 \\
\rho_{\rm T_+S\Downarrow\Uparrow}  & \rho_{\rm T_+S\Downarrow\Downarrow} & 0 & 0& 0 & 0 & 0& 0\\
\rho_{\rm T_0S\Uparrow\Uparrow} &\rho_{\rm T_0S\Uparrow\Downarrow}  & 0& 0& 0 & 0 & 0& 0\\
\rho_{\rm T_0S\Downarrow\Uparrow} &\rho_{\rm T_0S\Downarrow\Downarrow}  & 0&  0&0 & 0& 0 &  0 \\
\rho_{\rm T_-S\Uparrow\Uparrow} &\rho_{\rm T_-S\Uparrow\Downarrow}  & 0& 0& 0 & 0 & 0& 0\\
\rho_{\rm T_-S\Downarrow\Uparrow} &\rho_{\rm T_-S\Downarrow\Downarrow}  & 0&  0&0 & 0& 0 &  0 \\
\end{pmatrix}\label{rho8}
\eeq
\end{widetext}
How do we quantify how coherent is an ensemble of RPs described by some density matrix $\rho$? In other words, how do we measure the "strength" of $\rho_{\rm ST}+\rho_{\rm TS}$ ?  We call this measure $p_{\rm coh}$ and we treat it as a probability measure taking values between 0 (zero S-T coherence) and 1 (maximum S-T coherence). 

However, there are a few complications. First, $\rho$ carries information about all sorts of coherences, e.g. nuclear spin coherences. But we are interested in quantifying the coherence between the electron singlet and triplet subspace, irrespective of the nuclear spin state. For example, consider the two-nuclear-spin radical-pair in the state $\ket{\rm S}\otimes{1\over \sqrt{2}}(\ket{\Uparrow\Uparrow}+\ket{\Downarrow\Downarrow})$. This state contains nuclear spin coherence and, in fact, also nuclear spin entanglement, however it is an eigenstate of $\qs$, and hence has zero S-T coherence. We need a measure quantifying three types of singlet-triplet coherence that can in general exist, S-T$_+$, S-T$_0$ and S-T$_-$. In \cite{pre2014} we defined
\beq
C(\rho)=\sum_{j=0,\pm}\sqrt{\tr\{\rho_{\rm ST}|T_j\rangle\langle T_j|\rho_{\rm TS}\}}\label{crho},
\eeq
where $|{\rm T}_j\rangle\langle {\rm T}_j|$ is the projector\footnote{We did not explicitly include the tensor product with unit matrices for the nuclear spin degrees of freedom in defining the projectors $|{\rm T}_j\rangle\langle {\rm T}_j|$, which really are $d\times d$ matrices operating in the $d$-dimensional space of the radical-pair under consideration.} to one of the three triplet subspaces. This definition is motivated by considering the most general pure state of a radical-pair, written as $|\psi\rangle=\alpha|{\rm S}\rangle\otimes|\chi_{\rm S}\rangle+\sum_{j=0,\pm}\beta_j|{\rm T}_j\rangle\otimes|\chi_j\rangle$, where $\alpha$, $\beta_+$, $\beta_0$ and $\beta_-$ are complex amplitudes, and $|\chi_{\rm S}\rangle$, $|\chi_{+}\rangle$, $|\chi_{0}\rangle$ and $|\chi_{-}\rangle$ are arbitrary and normalized spin states of multiple nuclei. It is $C(\ket{\psi}\bra{\psi})=|\alpha\beta_1|+|\alpha\beta_0|+|\alpha\beta_{-1}|$, hence indeed $C(\rho)$ describes the combined "strength" of all three coherences. For the example states given in \eqref{rho4} and \eqref{rho8} it would be $C(\rho)=|\rho_{\rm ST_+}|+|\rho_{\rm ST_0}|+|\rho_{\rm ST_-}|$ and $C(\rho)=\sum_{j=0,\pm}\sqrt{|\rho_{\rm ST_j\Uparrow\Uparrow}|^2+|\rho_{\rm ST_j\Uparrow\Downarrow}|^2+|\rho_{\rm ST_j\Downarrow\Uparrow}|^2+|\rho_{\rm ST_j\Downarrow\Downarrow}|^2}$, respectively.

A second complication has to do with the normalization of $\rho$. Since $C(\rho)$ is linear in the matrix elements of $\rho$, scaling as $\tr\{\rho\}$, which is a decaying function of time due to charge recombination, $C(\rho)$ is also a decaying function of time. Therefore $C(\rho)$ has to be normalized by $\tr\{\rho\}$ in order to define the following probability measure.

To extract such a measure from $C(\rho)$, we must normalize $C(\rho)$ by its maximum. In \cite{pre2014} we calculated the maximum of $C(\rho)$ when only the Hamiltonian evolution is taken into account, i.e. ${\rm max}\{C(\tilde{\rho})\}$, where $d\tilde{\rho}/dt=-i[{\cal H},\tilde{\rho}]$. So in \cite{pre2014} we defined
\beq
p_{\rm coh}={1\over \tr\{\rho\}}{{C(\rho)}\over {{\rm max}\{C(\tilde{\rho})\}}}\label{pcoh1}
\eeq
In retrospect, this definition of the normalization is not satisfactory, because if one is given a density matrix $\rho$ describing an RP ensemble without any reference to ${\cal H}$, the measure $p_{\rm coh}$ cannot be calculated. Surely, in most calculations where one propagates the density matrix, the evolution is driven by a known Hamiltonian. But in principle it would be desirable to have a measure $p_{\rm coh}$ defined by just providing the density matrix $\rho$. Another way to normalize $C(\rho)$ based solely on $\rho$ is derived from the most general pure state $\ket{\psi}$ given before. We define the maximally S-T coherent pure state as the state having equal amplitudes in all four terms of $|\psi\rangle$, $|\alpha|=|\beta_+|=|\beta_0|=|\beta_-|$. Due to normalization it follows that $|\alpha|=|\beta_+|=|\beta_0|=|\beta_-|=1/2$ and hence $C(|\psi\rangle\langle\psi |)=3/4$. 
We can thus define 
\beq
p_{\rm coh}={4\over 3}{{C(\rho)}\over {\tr\{\rho\}}}\label{pcoh2}
\eeq
A few comments are in order: (i) the definition of $C(\rho)$ has the proper (linear) scaling with the density matrix elements to qualify \cite{plenio_coherence} for a proper measure of coherence. (ii) we defined the pure state of maximum coherence as in \cite{plenio_coherence}, where it is stated that in a space of dimension $N$ spanned by the vectors $|j\rangle$, where $j=1,2,...,N$, the state of maximum coherence is ${1\over\sqrt{N}}\sum_{j=1}^{N}|j\rangle$. (iii) in idealized theoretical scenarios, e.g. when one considers a 2-dimensional radical-pair spanned by just two states, $|{\rm S}\rangle$ and $|{\rm T}_0\rangle$, the normalization constant 3/4 changes accordingly, i.e. it would become 1/2. (iv) when we first introduced the need for an S-T coherence measure \cite{pre2011} we defined $p_{\rm coh}$ in a way that it scales quadratically with the matrix elements of $\rho_{\rm ST}+\rho_{\rm TS}$, and this is not an acceptable measure according to \cite{plenio_coherence}. The definition \eqref{crho} alleviates this problem. (v) the different normalizations in \eqref{pcoh1} and \eqref{pcoh2} produce acceptable numerical differences, but a more thorough study is required to fully understand the normalization of $C(\rho)$.
\subsubsection{Coherence Distillation}
There is one more step before proceeding with the theory of quantum retrodiction. As mentioned previously, the general RP density matrix can be written as 
$\rho=\rho_{\rm SS}+\rho_{\rm TT}+\rho_{\rm ST}+\rho_{\rm TS}$. Having defined the coherence measure $p_{\rm coh}(\rho)$, we can introduce two new matrices, the maximally incoherent and maximally coherent version of $\rho$, denoted by $\rho_{\rm incoh}$ and $\rho_{\rm coh}$, respectively, and given by
\begin{align}
\rho_{\rm incoh}&=\rho_{\rm SS}+\rho_{\rm TT}\label{rincoh}\\
\rho_{\rm coh}&=\rho_{\rm SS}+\rho_{\rm TT}+{1\over p_{\rm coh}}(\rho_{\rm ST}+\rho_{\rm TS})\label{rcoh}
\end{align}
While the interpretation $\rho_{\rm incoh}$ is obvious, the interpretation of $\rho_{\rm coh}$ is more subtle. Reminiscent of the concept of entanglement distillation \cite{dist1,dist2}, $\rho_{\rm coh}$ alludes to the coherence-distilled version of $\rho$, since $p_{\rm coh}(\rho_{\rm coh})=1$. To give a simple example, consider like in Section 3.7.1 a mixture of $n$ radical-pairs in the state $|\psi_1\rangle=(|{\rm S}\rangle+|{\rm T}_0\rangle)/\sqrt{2}$ and $m$ radical-pairs in the singlet state $|\psi_2\rangle=|{\rm S}\rangle$. It is $p_{\rm coh}=n/(n+m)$, so $p_{\rm coh}$ should be the probability that picking an RP out of the ensemble it will have maximum S-T coherence. In general, using \eqref{rincoh} and \eqref{rcoh} we can write any density matrix $\rho$ as $\rho=(1-p_{\rm coh})\rho_{\rm incoh}+p_{\rm coh}\rho_{\rm coh}$.
\subsubsection{Quantum retrodiction}
\begin{figure}
\begin{center}
\includegraphics[width=7 cm]{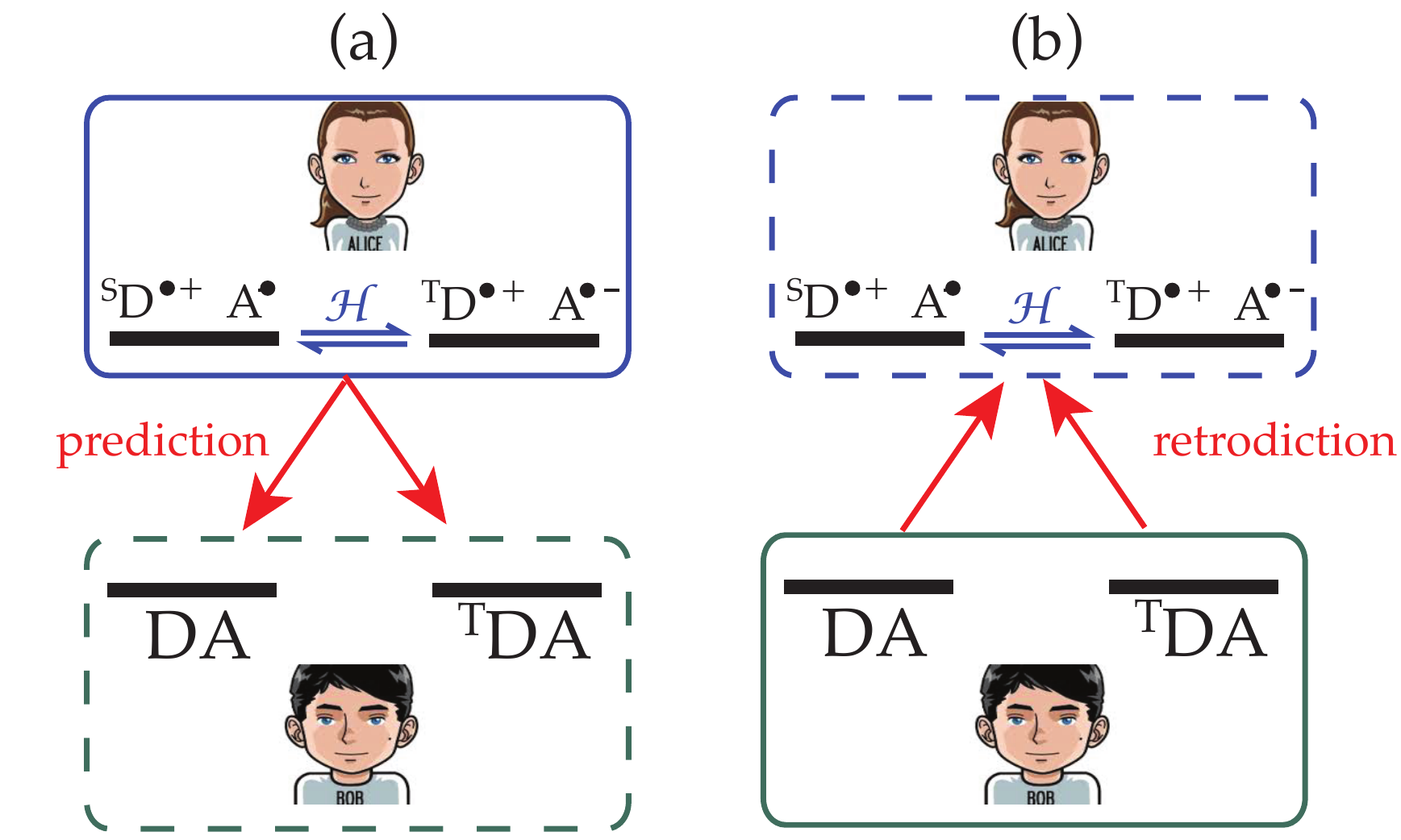}
\caption{Quantum communication model of the radical-pair mechanism. Alice holds the quantum state of the radical-pair. Bob only has access the the recombination products, which are either singlet or triplet. In the predictive direction (a) we know the radical-pair state and calculate the probabilities of Bob's measurement outcomes. In the retrodictive direction (b) we have access to a specific recombination product and try to estimate the pre-recombination radical-pair state.}
\label{qc}
\end{center}
\end{figure}

In Fig. \ref{qc} we show that RP reactions can be viewed as a biochemical realization of quantum communication. Indeed, the RP spin state might be considered as the quantum state held by Alice, the sender. Bob just detects singlet or triplet neutral products. From these he might infer the information stored in Alice's states, for example the value of the applied magnetic field. 
Quantum retrodiction probabilistically answers the question of what was the pre-recombination state of a radical-pair, the neutral product of which is detected by Bob. 
The relevant formalism \cite{retro3,retro4} is used in \cite{pre2014} to calculate the conditional probabilities that upon detecting a singlet (triplet) product, the pre-recombination state of the radical-pair was maximally coherent or maximally incoherent, $P({\rm incoh}|{\rm S})=P({\rm incoh}|{\rm T})=1-p_{\rm coh}$, and $P({\rm coh}|{\rm S})=P({\rm coh}|{\rm T})=p_{\rm coh}$. To arrive at the reaction terms, we need to subtract the estimated pre-recombination state of the recombined radical-pair from the density matrix of the surviving RPs. This crucially depends on S-T coherence \cite{pre2014}. For a maximally coherent density matrix it is
\begin{equation*}
\delta\rho_{\rm coh}^{\rm 1S}=\delta\rho_{\rm coh}^{\rm 1T}=-{\rho\over {\tr\{\rho\}}},
\end{equation*}
whereas for a maximally incoherent density matrix it is 
\begin{equation*}
\delta\rho_{\rm incoh}^{\rm 1S}=-{{{\rm Q_S}\rho{\rm Q_S}}\over {\tr\{\rho {\rm Q_S}\}}},~~
\delta\rho_{\rm incoh}^{\rm 1T}=-{{{\rm Q_T}\rho{\rm Q_T}}\over {\tr\{\rho {\rm Q_T}\}}}
\end{equation*}
Stated in words, these rules mean that in the extreme of maximum coherence we have to subtract the full pre-recombination state from $\rho$, whereas in the opposite extreme we subtract either a singlet or a triplet RP. The previous update rules hold for a single molecule. During $dt$ we will detect $dr_{\rm S}$ singlet and $dr_{\rm T}$ triplet products, hence the reaction terms read
\beq
{{d\rho}\over {dt}}\Big |_{\rm recomb}=dr_{\rm S}\delta\rho^{\rm 1S}+dr_{\rm T}\delta\rho^{\rm 1T},\label{recomb}
\eeq
where 
\begin{align}
\delta\rho^{\rm 1S}&=P({\rm incoh}|{\rm S})\delta\rho_{\rm incoh}^{\rm 1S}+P({\rm coh}|{\rm S})\delta\rho_{\rm coh}^{\rm 1S}\nonumber\\
\delta\rho^{\rm 1T}&=P({\rm incoh}|{\rm T})\delta\rho_{\rm incoh}^{\rm 1T}+P({\rm coh}|{\rm T})\delta\rho_{\rm coh}^{\rm 1T}\nonumber
\end{align}
\subsection{Radical-pair master equation}
As already mentioned, singlet-triplet dephasing and spin-dependent recombination are two physical processes running simultaneously and independently, while stemming from the same interaction Hamiltonian between radical-pair and vibrational reservoir. Using \eqref{MEnr} and \eqref{recomb}, we thus arrive at the sought after master equation
\begin{align}
{{d\rho}\over {dt}}&={{d\rho}\over {dt}}\Big |_{\rm decoh}+{{d\rho}\over {dt}}\Big |_{\rm recomb}\nonumber\\&=-i[{\cal H},\rho]+{\cal D}\llbracket\rho\rrbracket+{\cal R}_{K}\llbracket\rho\rrbracket\label{ME}
\end{align}
where ${\cal R}_{K}\llbracket\rho\rrbracket$ is the reaction super-operator 
\begin{align}
{\cal R}_{K}\llbracket\rho\rrbracket=&-(1-p_{\rm coh})\big(\ks\qs\rho\qs+\kt\qt\rho\qt\big)\nonumber\\&-p_{\rm coh}{{dr_{\rm S}+dr_{\rm T}}\over {dt}}{\rho_{\rm coh}\over {\tr\{\rho\}}}\label{RK}
\end{align}
The term $-i[{\cal H},\rho]$ generates S-T coherence, which is dissipated by ${\cal D}\llbracket\rho\rrbracket$, while ${\cal R}_{K}\llbracket\rho\rrbracket$ accounts for the state-dependent recombination.  As expected, the normalization of $\rho$ decreases by the neutral products appearing in the time interval $dt$, i.e. from \eqref{ME} we find $d{\rm Tr}\{\rho\}=-dr_{\rm S}-dr_{\rm T}$. Given the particular Hamiltonian ${\cal H}$ and recombination rates $\ks$ and $\kt$, and using the definition of $p_{\rm coh}$, it is straightforward to propagate $\rho$ starting from a specific initial condition.
\subsection{Examination of special cases}
\subsubsection{Recombination without S-T mixing}
Consider an imaginary RP having a non-reactive triplet state and no S-T mixing, i.e. take $\kt=0$ and ${\cal H}=0$.

(i) singlet initial state, $\rho_0=\qs/\tr\{\qs\}$. We expect the RP population to decay exponentially at a rate $\ks$, the surviving RP state remaining the same as at $t=0$. Indeed, it is ${\cal D}\llbracket\rho_0\rrbracket=0$ and $p_{\rm coh}=0$. Plugging these and ${\cal H}=0$, $\kt=0$ into \eqref{ME} leads to $d\rho/dt=-\ks\rho$, so $\rho_t=e^{-\ks t}\qs/\tr\{\qs\}$, as expected.

(ii) maximally incoherent S-T mixture\footnote{We refer to an {\it improper} mixture, see Section 7}, i.e. if $\rho_0=\alpha\qs/\tr\{\qs\}+\beta\qt/\tr\{\qt\}$, with $\alpha,\beta$ real and $\alpha+\beta=1$. We find that $d\rho/dt=-\ks\qs\rho\qs$, so that $\rho_t=\alpha e^{-\ks t}\qs/\tr\{\qs\}+\beta\qt/\tr\{\qt\}$, as expected.

(iii) maximally coherent superposition $|\psi_0\rangle=(|{\rm S}\rangle+|{\rm T}_0\rangle)/\sqrt{2}$. Now the master equation cannot be solved analytically, since at $t=0$ we have a maximally coherent state with $p_{\rm coh}=1$, but right in the following interval $dt$ the S-T dephasing super-operator ${\cal D}\llbracket\rho\rrbracket$ will produce a partially coherent state having $p_{\rm coh}<1$, hence all terms of ${\cal R}_{K}\llbracket\rho\rrbracket$ will start contributing. However, as a preamble to quantum trajectories to be discussed in Section 5, we note four possibilities for state evolution during the interval $dt$: (1) singlet projection, with probability $dp_{\rm S}=\ks dt/4$, (2) triplet projection with probability $dp_{\rm T}=\ks dt/4$, (3) singlet recombination with probability $dr_{\rm S}=\ks dt/2$, and  (4) no evolution, i.e. the state remains $|\psi_0\rangle$, with probability $1-dp_{\rm S}-dp_{\rm T}-dr_{\rm S}=1-\ks dt$. Those RPs projected at some point to the non-reactive triplet state will remain there. Their total number as $t\rightarrow\infty$ is $\ks dt/4+(1-\ks dt)\ks dt/4+(1-\ks dt)^2\ks dt/4+...=1/4$. 

At first sight this is a weird result challenging our understanding of quantum measurements. Starting with $\ket{\psi_0}$, in which the probability for the RP to be in the triplet state is 1/2, we find that only 25\% of the RPs are locked in the non-reactive triplet state at the end of this reaction. The apparent contradiction with our intuition is due a cavalier phrasing in the previous argument. Indeed, if we prepare an RP ensemble in the state $|\psi_0\rangle$ and {\it immediately perform a global projective measurement on the ensemble}, we will find 50\% of the RPs in $\ket{\rm S}$ and 50\% in $\ket{\rm T_0}$. But the radical-pair reaction is not a global projective measurement performed at once on all RPs. Instead, it proceeds as a continuous intra-molecule measurement leading to the already discussed random projections. These enhance the singlet yield, since those RPs projected to the singlet state will definitely recombine at some point.
\subsubsection{Equal recombination rates}
The case $\ks=\kt\equiv k$ is especially simple, since the reaction is spin-independent, so ${\cal R}_{K}\llbracket\rho\rrbracket$ should not have any effect on the RP state. Indeed, now it is ${\cal R}_{K}\llbracket\rho\rrbracket=-k\rho$, hence $d\rho/dt=-i[{\cal H},\rho]-k(\rho\qs+\qs\rho-2\qs\rho\qs)-k\rho$. Defining $\rho=e^{-kt}R$, it follows that $dR/dt=-i[{\cal H},R]-k(R\qs+\qs R-2\qs R\qs)$, i.e. the state of the RPs evolves just as if they never recombine, of course with their total population decaying exponentially. If we now consider the case ${\cal H}=0$ and write $R(t)=R_{\rm SS}(t)+R_{\rm TT}(t)+R_{\rm ST}(t)+R_{\rm TS}(t)$ we easily find 
$R(t)=R_{\rm SS}(0)+R_{\rm TT}(0)+e^{-kt}\big(R_{\rm ST}(0)+R_{\rm TS}(0)\big)$, meaning that the initially present coherence $\rho_{\rm ST}(0)+\rho_{\rm TS}(0)$ decays at the rate $2k$, while the population decays a the rate $k$. That is,
coherence is damped due to the decaying population, but also due to the intrinsic S-T dephasing process.
\section{Haberkorn's master equation}
If the master equation \eqref{ME} we arrived at is a deeper theory capturing the quantum dynamics of RP reactions, then Haberkorn's master equation must be some sort of limiting case of our theory. It is readily seen that Haberkorn's master equation \eqref{trad} can be recast in the form $d\rho/dt=-i[{\cal H},\rho]+{\cal D}\llbracket\rho\rrbracket+{\cal R}_{H}\llbracket\rho\rrbracket$, where ${\cal R}_{H}\llbracket\rho\rrbracket$ is Haberkorn's reaction operator:
\beq
{\cal R}_{H}\llbracket\rho\rrbracket=-\ks\qs\rho\qs-\kt\qt\rho\qt\label{RH}
\eeq
If by hand we force $p_{\rm coh}=0$ in our reaction operator ${\cal R}_{K}\llbracket\rho\rrbracket$, we exactly retrieve ${\cal R}_{H}\llbracket\rho\rrbracket$. In other words, the S-T dephasing process ${\cal D}\llbracket\rho\rrbracket$ is latently built into Haberkorn's master equation, however the reaction terms \eqref{RH} skew the state evolution of RPs by retrodicting the pre-recombination state always assuming zero S-T coherence.

In his 1976 paper Haberkorn had qualitatively examined various forms of the master equation, one being the decoherence master equation \eqref{MEnr}. However, he correctly dismissed it since it is trace-preserving and hence "corresponds to spin relaxation without reaction". We now understand that S-T dephasing is just one aspect of the dynamics running simultaneously with the other, the RP recombination. In any case, Haberkorn's 1976 paper is a tribute to the power of simple physical arguments leading to important insights.
\subsection{When is the traditional master equation an adequate theory?}
The traditional master equation \eqref{trad} has been used in most theoretical considerations in the field of spin chemistry since the late 1960s, adequately accounting for many experiments. So a natural question to ask is: why do we need a new theory when the one at hand, even if not fundamentally sound, appears to be good enough? Towards the answer we first remind the reader of the early era of atomic physics, when atom-light interactions were successfully described by Einstein's rate equations, which just considered the transfer of atomic populations between atomic states, the transfer being triggered by (at the time incoherent) light sources such as lamps. After the invention of the laser, the field of coherent atomic interactions exploded, since atomic coherences could then be excited and probed. Einstein's rate equations were not sufficient to describe experiments, which required the introduction of the atomic density matrix, i.e. the coupled time evolution of atomic populations and coherences. The parameter deciding which approach is the most suitable is the ratio of the Rabi frequency of the exciting light to the relaxation rate, which in this case is the spontaneous decay rate. 

In spin chemistry, many experiments so far have been apparently dominated by spin relaxation, which among other things, also damps S-T coherence quantified by $p_{\rm coh}$. If these relaxation phenomena push $p_{\rm coh}$ towards zero much faster than the fundamental decoherence mechanism we have considered, then our master equation quickly converges to the traditional theory. This will be further quantified in the following subsection.

We can thus arrive at the conclusion that, notwithstanding fortuitous agreement with experiments brought about by relaxing environments, understanding spin chemistry experiments at the fundamental level is not possible within the traditional theory. Moreover, without a fundamental theory it is neither possible to predict the full range of physical effects that can be {\it in principle} observed, nor design new experiments and explore new fronts of the field. For example, understanding the avian compass mechanism at the quantum level, e.g. addressing the fundamental heading error of the compass and its magnetic sensitivity \cite{cpl2012_1} is a promising venue of biochemical quantum metrology, which however cannot be conclusively explored without the fundamental theory of RPM. This is because even if radical-pairs participating in the natural avian compass are plagued by relaxation, future biomimetic sensors need not be.  

Finally, there are cases where any agreement with experiments is deluding, i.e. Haberkorn's incorrect reaction super-operator forces other system parameters (like Hamiltonian couplings) to take on incorrect values in order for the theory to match the data. As will be elaborated in Section 5.2, such an example are the spin dynamics in CIDNP, which cannot be understood within Haberkorn's approach, since the lifetimes of the involved RPs are too short for relaxation to set in. 
\subsubsection{Additional relaxation processes mask the underlying quantum dynamics of radical-pairs}
A general relaxation process can be described by a set of Kraus operators $K_n$, satisfying $\sum_{n}K_{n}^{\dagger}K_{n}=\mathbbmtt{1}$ and transforming the density matrix according to $\rho\rightarrow\rho'=\sum_{n}K_{n}\rho K_{n}^{\dagger}$, i.e. the master equation \eqref{ME} will be augmented with the term ${\cal K}\llbracket\rho\rrbracket=\Big(\sum_{n}K_{n}\rho K_{n}^{\dagger}-\rho\Big)/dt$. Note that ${\cal D}\llbracket\rho\rrbracket$ is a special case of ${\cal K}\llbracket\rho\rrbracket$ resulting from the following three Kraus operators: $K_1=\sqrt{1-\lambda}{\rm {\rm Q_S}}$, $K_2=\sqrt{1-\lambda}{\rm Q_T}$ and $K_3=\sqrt{\lambda}\mathbbmtt{1}$, with $\lambda=1-(\ks+\kt)dt/2$.

The simplest way to make our point is to assume that besides the fundamental S-T dephasing process, we have an independent relaxation mechanism of rate $\beta$, being described by the same Kraus operators $K_1$, $K_2$ and $K_3$. Defining\footnote{Note the different notation from the coherence measure $C(\rho)$, which takes as input an operator and produces a number. In contrast, the calligraphic ${\cal C}\llbracket .\rrbracket$ takes as input an operator and produces another operator.}  the super-operator ${\cal C}\llbracket\rho\rrbracket=\qs\rho\qt+\qt\rho\qs$, we use \eqref{ME} to find that ${\cal C}\llbracket\rho\rrbracket$ obeys the equation
\beq
{{d{\cal C}\llbracket\rho\rrbracket}\over{dt}}=-i{\cal C}\llbracket[{\cal H},\rho]\rrbracket-\Gamma_{c}{\cal C}\llbracket\rho\rrbracket\label{drcdt},
\eeq
where $\Gamma_{c}=\beta+k_{\rm S}\Big({1\over 2}+\langle{\rm \tilde{Q}_S}\rangle\Big)+k_{\rm T}\Big({1\over 2}+\langle{\rm \tilde{Q}_T}\rangle\Big)$, 
and we defined $\langle{\rm \tilde{Q}_x}\rangle=\tr\{\rho{\rm Q}_x\}/\tr\{\rho\}$ with $x={\rm S,T}$. Taking the trace of \eqref{ME}, we find that ${{d\tr\{\rho\}}/{dt}}=-\kappa\tr\{\rho\}$, where $\kappa=k_{\rm S}\langle{\rm \tilde{Q}_S}\rangle+k_{\rm T}\langle{\rm \tilde{Q}_T}\rangle$. We finally define the "genuine" S-T decoherence rate as $\gamma_{c}=\Gamma_{c}-\kappa$. This describes the decay of S-T coherence due to all effects other than the changing normalization of $\rho$. For the considered example it is $\gamma_{c}=(\ks+\kt)/2+\beta$. It is obvious that if $\beta\gg \Omega,(\ks+\kt)/2$, where $\Omega$ is the rate at which S-T coherence is generated by the Hamiltonian, ${\cal C}\llbracket\rho\rrbracket$ will quickly (compared to the reaction time) decay to zero, and thus $p_{\rm coh}(\rho)\approx 0$. Hence the traditional theory will provide a consistent description of the dynamics. 
\subsection{Haberkorn's reaction operator contradicts the physics of quantum state reconstruction}
The form of Haberkorn's reaction operator \eqref{RH} can be simply phrased as follows: singlet products originate from singlet radical-pairs, and triplet products originate from triplet radical-pairs. This reasoning obliterates the very concepts of quantum superposition and quantum measurement and is unphysical on many grounds, one related to the well understood quantum state reconstruction. Suppose we are provided with a number of qubits all prepared in the state $|\psi\rangle=(|0\rangle+|1\rangle)/\sqrt{2}$, which however is not disclosed to us. Suppose then that we perform a measurement of each qubit either in the basis B1=$\{|0\rangle,|1\rangle\}$ or in the basis B2=$\{{{|0\rangle+|1\rangle}\over \sqrt{2}},{{|0\rangle-|1\rangle}\over \sqrt{2}}\}$. Measuring in B1, the measurement outcomes will be 0 or 1 and we should conclude that the preparation state was $|0\rangle$ or $|1\rangle$. Measuring in B2, the outcome will always be 0, and we should conclude that the prepared state was $|\psi\rangle$. Repeating this many times we should conclude that we are provided with qubits in all three states $|\psi\rangle$, $|0\rangle$ and $|1\rangle$. In other words, based on the dictum of identifying the pre-measurement state with the post-measurement state, we would never be able to use the statistical-interpretation of quantum measurements and reconstruct the original state preparation. 
\subsection{Haberkorn's master equation coherently generates entanglement starting out with none}
We will again consider an RP ensemble initially prepared in the state $|\psi_0\rangle=(|{\rm S}\rangle+|{\rm T}_0\rangle)/\sqrt{2}$, not undergoing any S-T mixing (${\cal H}=0$) and having $\kt=0$. Haberkorn's theory predicts that the RP state remains pure at all times, with the ensemble consisting at time $t$ of $(1+e^{-\ks t})/2$ radical-pairs in the state 
\beq
|\psi_t\rangle={1\over \sqrt{1+e^{-\ks t}}}\Big(e^{-{{\ks t}\over 2}}|{\rm S}\rangle+|{\rm T}_0\rangle\Big)\label{psi1}
\eeq
At $t=0$ it is $|\psi_0\rangle=\ket{\uparrow\downarrow}$, which is not entangled. However, it is seen from \eqref{psi1} that {\it without any coherent operation} on the radical-pairs, they spontaneously acquire a non-zero entanglement while {\it coherently} evolving to the maximally entangled and non-reactive triplet state $|{\rm T}_0\rangle$. This is unphysical. The caveat, as will be explained in detail in Section 6.2, is that an irreversible reaction causing population leakage is wrongly understood to coherently operate on the amplitudes of the RP quantum state. In contrast, in our approach an RP in the state $|\psi_0\rangle$ is (i) either projected to $|{\rm S}\rangle$, or (ii) projected to $|{\rm T}_0\rangle$, or (iii) recombines to a singlet neutral product, or (iv) stays put at $|\psi_0\rangle$. The probabilities of (i) and (ii) are the same (see Section 3.8.1), so together they represent a balanced mixture of maximally entangled states, the mixture carrying zero entanglement. Event (iii) produces net entanglement, but it does so by an incoherent operation, related to measurement-induced entanglement, to be addressed in more detail elsewhere.
\section{Haberkorn's master equation is not consistent with the quantum-trajectory picture}
In Section 3.5 we discussed the interpretation of \eqref{MEnr} in terms of single-molecule quantum trajectories involving three possible events, (i) singlet projection with probability $dp_{\rm S}$, (ii) triplet projection with probability $dp_{\rm T}$, and (iii) Hamiltonian evolution with probability $1-dp_{\rm S}-dp_{\rm T}$. 

However, the full quantum dynamics include both S-T dephasing and recombination. To account for the latter, we have to augment the quantum trajectories, which are now formed by 5 possible events taking place within $dt$: (E1) singlet projection with probability $dp_{\rm S}$, (E2) triplet projection with probability $dp_{\rm T}$, (E3) singlet recombination with probability $dr_{\rm S}$, (E4) triplet recombination with probability $dr_{\rm T}$, and (E5) Hamiltonian evolution with probability $1-dp_{\rm S}-dp_{\rm T}-dr_{\rm S}-dr_{\rm T}$. Does the average of many quantum trajectories reproduce our master equation? Moreover, which are the quantum trajectories in Haberkorn's approach, and do they reproduce Haberkorn's master equation? 
\subsection{Quantum trajectories in Haberkorn's approach}
The concept of quantum trajectories, well known in quantum optics \cite{wiseman,wiseman_book}, was only recently introduced \cite{pre2014} in spin chemistry. In any case, in \cite{cpl2015} we presented a significant constraint to be met when one attempts a quantum-trajectory analysis of Haberkorn's master equation. 

The intuitive understanding in spin chemistry was that RPs evolve {\it unitarily} under the action of ${\cal H}$ until the instant they recombine into neutral reaction products. We have shown in \cite{cpl2015} that this physical picture {\it must} be advocated if Haberkorn's theory is to be consistent in the special case $\ks=\kt$. First of all, what is so special about this case? The rates $\ks$ and $\kt$ are RP-specific parameters entering the master equation, which obviously must be valid for all radical-pairs, those for which $\ks=\kt$ and those for which $\ks\neq\kt$. In fact the latter are abundant in photosynthetic reaction centers, as will be presented in the following section. Moreover, quantum dynamics are non-trivial in this case of asymmetric recombination rates. In contrast, RP quantum dynamics simplify a lot when $\ks=\kt\equiv k$, where RP population decays exponentially at a rate $k$, {\it without the decay affecting the state of the surviving RPs}. Haberkorn's master equation \eqref{trad} now becomes:
\beq
d\rho/dt=-i[{\cal H},\rho]-k\rho\label{drdtk}
\eeq
Defining a new density matrix $R$ by $\rho=e^{-kt}R$, it is $dR/dt=-i[{\cal H},R]$. Thus, apart from an exponential decay of the normalization $\tr\{\rho\}$, the density matrix $\rho$ evolves unitarily. Switching to quantum trajectories, we can exactly retrieve \eqref{drdtk} if we assume that a single RP faces just three possibilities along its state evolution during $dt$: (i) singlet recombination with probability $dr_{\rm S}$, (ii) triplet recombination with probability $dr_{\rm T}$, and (iii) Hamiltonian evolution with probability $1-dr_{\rm S}-dr_{\rm T}$. Indeed, the single-molecule state will be $\rho/\tr\{\rho\}$, hence averaging (i)-(iii) we get:
\begin{align}
\rho+d\rho&=dr_{\rm S}\Big(\rho-{\rho\over {\tr\{\rho\}}}\Big)+dr_{\rm T}\Big(\rho-{\rho\over {\tr\{\rho\}}}\Big)\nonumber\\&+(1-dr_{\rm S}-dr_{\rm T})\Big(\rho-i[{\cal H},\rho]dt\Big)\label{avg}
\end{align}
In the above equation we update $\rho$ by removing a single-molecule density matrix, weighing the result by the respective probability for singlet (first term) and triplet (second term) recombination, then weighing in the Hamiltonian evolution occuring if none of these two possibilities materialize. Taking into account that $dr_{\rm S}=kdt\tr\{\rho\qs\}$ and $dr_{\rm T}=kdt\tr\{\rho\qt\}$, and further that $dr_{\rm S}+dr_{\rm T}=kdt\tr\{\rho\}$ due to the completeness relation $\qs+\qt=\mathbbmtt{1}$, equation \eqref{avg} readily leads to \eqref{drdtk}. 
\begin{figure}
\begin{center}
\includegraphics[width=8cm]{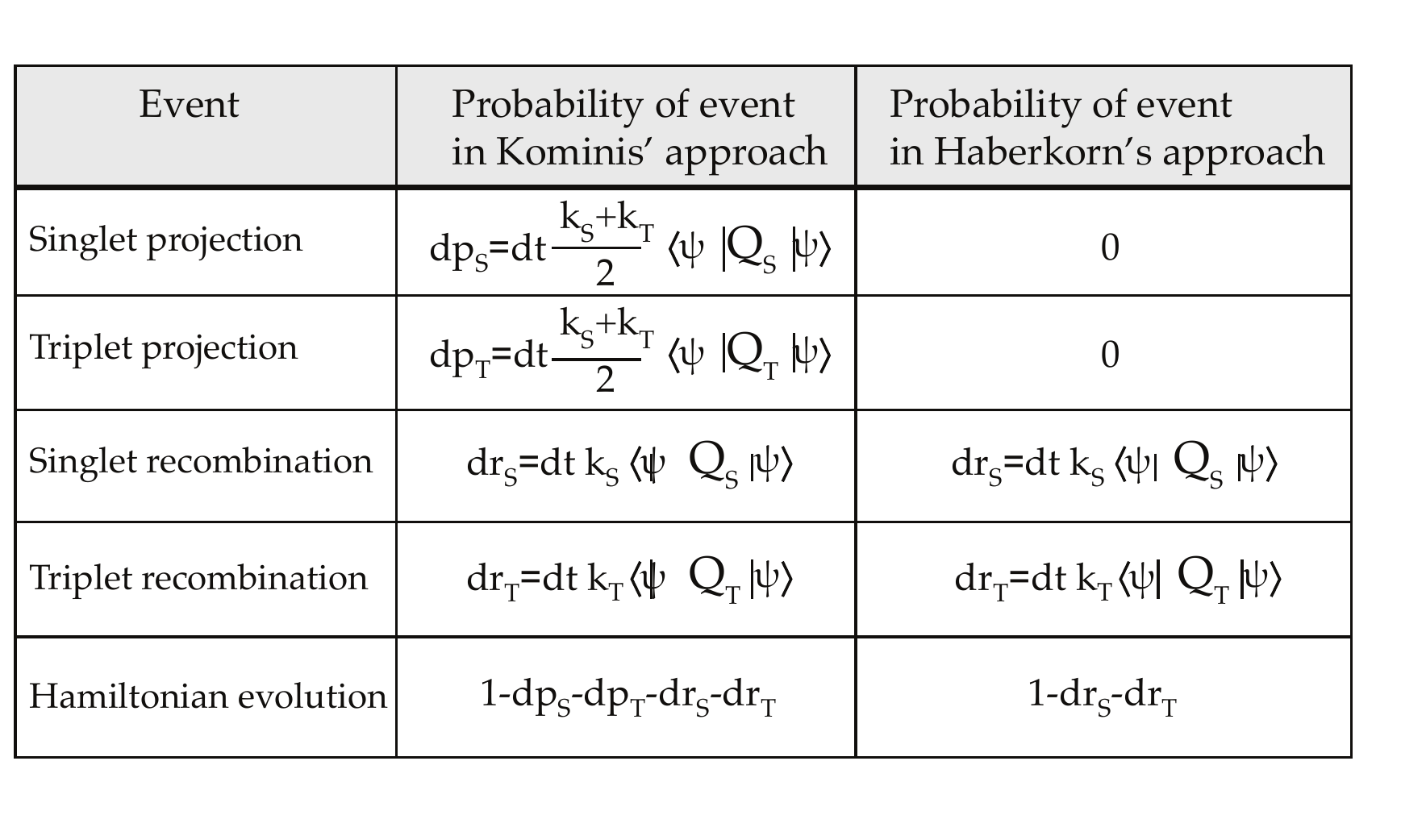}
\caption{Possible events forming quantum trajectories in the approach of Kominis and Haberkorn.}
\label{qtrajKH}
\end{center}
\end{figure}

As promised, we have arrived at a significant constraint limiting our freedom to define Haberkorn's quantum trajectories, namely that in the special case $\ks=\kt$, Haberkorn's master equation is consistent with quantum trajectories in which a radical-pair just evolves unitarily until it recombines. At this point we have to forfeit our goal to conclusively explore what sort of quantum trajectories Haberkorn's proponents might stipulate in the general case $\ks\neq\kt$. Starting from the physical model of the radical-pair, depicted in Fig. \ref{rip_levels}, and pinning down the spin-selective interaction Hamiltonian ${\cal V}$ coupling the radical-pair with the vibrational reservoirs, one is unambiguously led to the singlet and triplet projections of our approach. Being unable to imagine what would be a first-principles derivation of some different sort of quantum trajectories, which would not disturb RP state evolution in case $\ks=\kt$, we make the working assumption that for any values of $\ks$ and $\kt$ the quantum trajectories Haberkorn's proponents would pick are the same as for the special case $\ks=\kt$. In Fig. \ref{qtrajKH} we summarize both pictures of quantum trajectories to be used in the following consistency check. 
\subsection{Chemically induced dynamic nuclear polarization}
As mentioned in the introduction of Section 2, CIDNP is a versatile tool to study photosynthetic reaction centers with significantly enhanced NMR signals resulting from the dynamics of spin transport taking place along with charge transport. It is beyond the scope of this review to address the rich field of CIDNP. An excellent recent review is \cite{matysik_review}. In general, nuclear spin dynamics in CIDNP are regulated by the radical-pair mechanism, and similarly with EPR \cite{mobius}, from the measured NMR spectra one can obtain rich structural information about the reaction centers, e.g. electronic structure of the photochemically active cofactors. We have recently shown \cite{cpl2015} that CIDNP is exquisitely sensitive to the quantum dynamics of radical-pair reaction kinetics, since it interestingly turns out that for a large class of bacterial and higher-plant reaction centers, the recombination rates of the participating donor-acceptor systems are highly asymmetric, $\kt\gg\ks$ \cite{matysik_PR}. In this regime non-trivial quantum effects in the state evolution driven by the reaction super-operator are most dominant.

Instead of comparing absolute quantitative predictions of Haberkorn's theory against our theory, in \cite{cpl2015} we used CIDNP to test the internal consistency of both theories. It is known \cite{wiseman,wiseman_book} that the predictions of the master equation must exactly coincide with the predictions resulting from the average of many single-molecule quantum trajectories. 
\begin{figure}
\begin{center}
\includegraphics[width=8 cm]{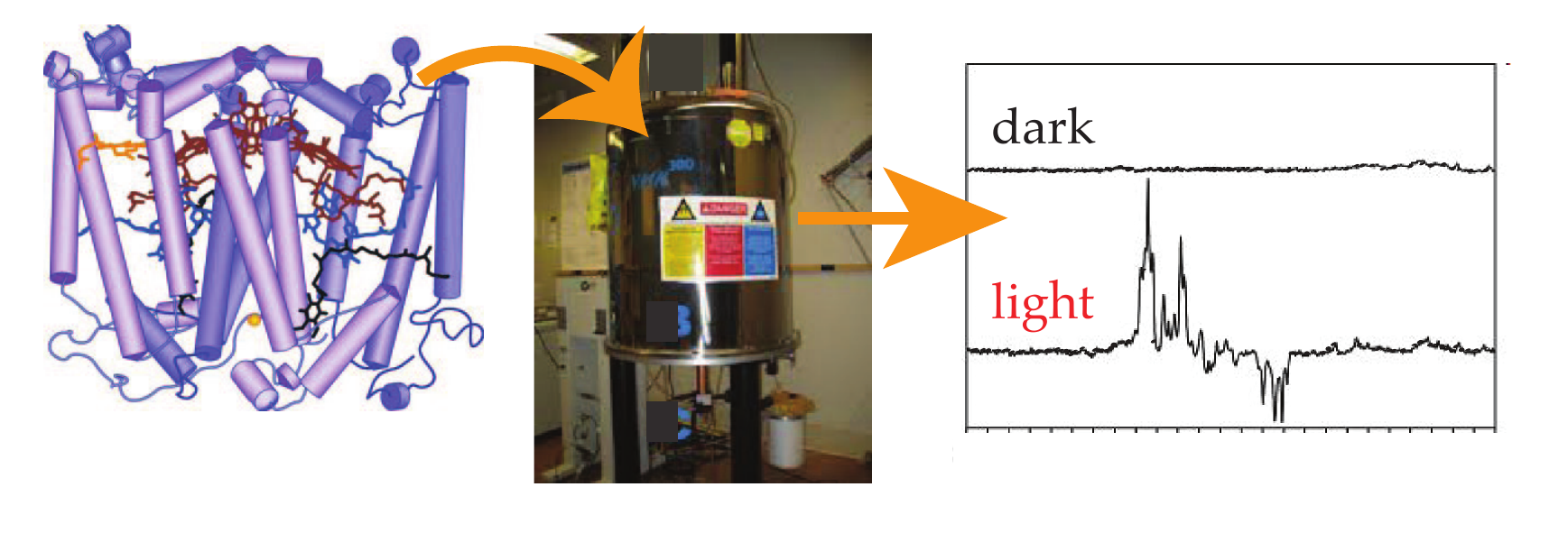}
\caption{Schematic of a solid-state photo-CIDNP experiment (the NMR signal was taken from http://ssnmr.lic.leidenuniv.nl/files/ssnmr/10-03.Matysik\%20CIDNP.pdf). A solid sample of reaction centers is placed in a high-field NMR magnet. Upon illumination, an NMR signal is observed, enhanced by several orders of magnitude above thermal equilibrium.}
\label{cidnp}
\end{center}
\end{figure}
The results of this test performed in \cite{cpl2015} demonstrate that while our approach is largely albeit not perfectly consistent, Haberkorn's approach is highly inconsistent, and hence is ruled out. We used a Hamiltonian of the same form considered in CINDP works like \cite{jeschke1998}, 
\beq
{\cal H}={{\Delta\omega}\over 2}s_{Az}-{{\Delta\omega}\over 2}s_{Dz}+\omega_{I}I_{z}+As_{Az}I_{z}+Bs_{Az}I_{x},\label{ham}
\eeq
where the relevant parameters are outlined in \cite{cpl2015}. Importantly, we use asymmetric rates $\kt\gg\ks$ pertinent to photosynthetic reaction centers \cite{matysik_review}. 
We calculate the nuclear spin deposited to the reaction products, $dI_z$, as a function of time \cite{JM}. When propagating the density matrix, the properly normalized density matrix of the singlet and triplet neutral products is\footnote{With the bar we denote spin states of the neutral products.} $\bar{\rho}_{\rm x}={\rm Q_x}\rho{\rm Q_x}/\tr\{\rho{\rm Q_x}\}$, while their number produced during $dt$ is $dr_{\rm x}=k_{\rm x} dt\tr\{\rho{\rm Q_x}\}$, with ${\rm x=S,T}$. Hence the total (singlet + triplet) ground-state nuclear spin accumulated during $dt$ is 
\begin{align}
dI_z&=dr_{\rm S}\tr\{I_z\bar{\rho}_{\rm S}\}+dr_{\rm T}\tr\{I_z\bar{\rho}_{\rm T}\}\nonumber\\&=dt\Big(\ks\tr\{I_z\qs\rho\qs\}+\kt\tr\{I_z\qt\rho\qt\}\Big)
\end{align}
When propagating quantum trajectories, the properly normalized state of the singlet and triplet reaction product is $|\bar{\psi}_{x}\rangle={\rm Q_x}|\psi\rangle/\sqrt{\langle\psi|{\rm Q_x}|\psi\rangle}$, and the nuclear spin deposited to the reaction product is $\langle\bar{\psi}_{\rm x}|I_{z}|\bar{\psi}_{\rm x}\rangle$ for a trajectory terminating with a x-recombination, where ${\rm x=S,T}$.
\begin{figure}
\begin{center}
\includegraphics[width=7 cm]{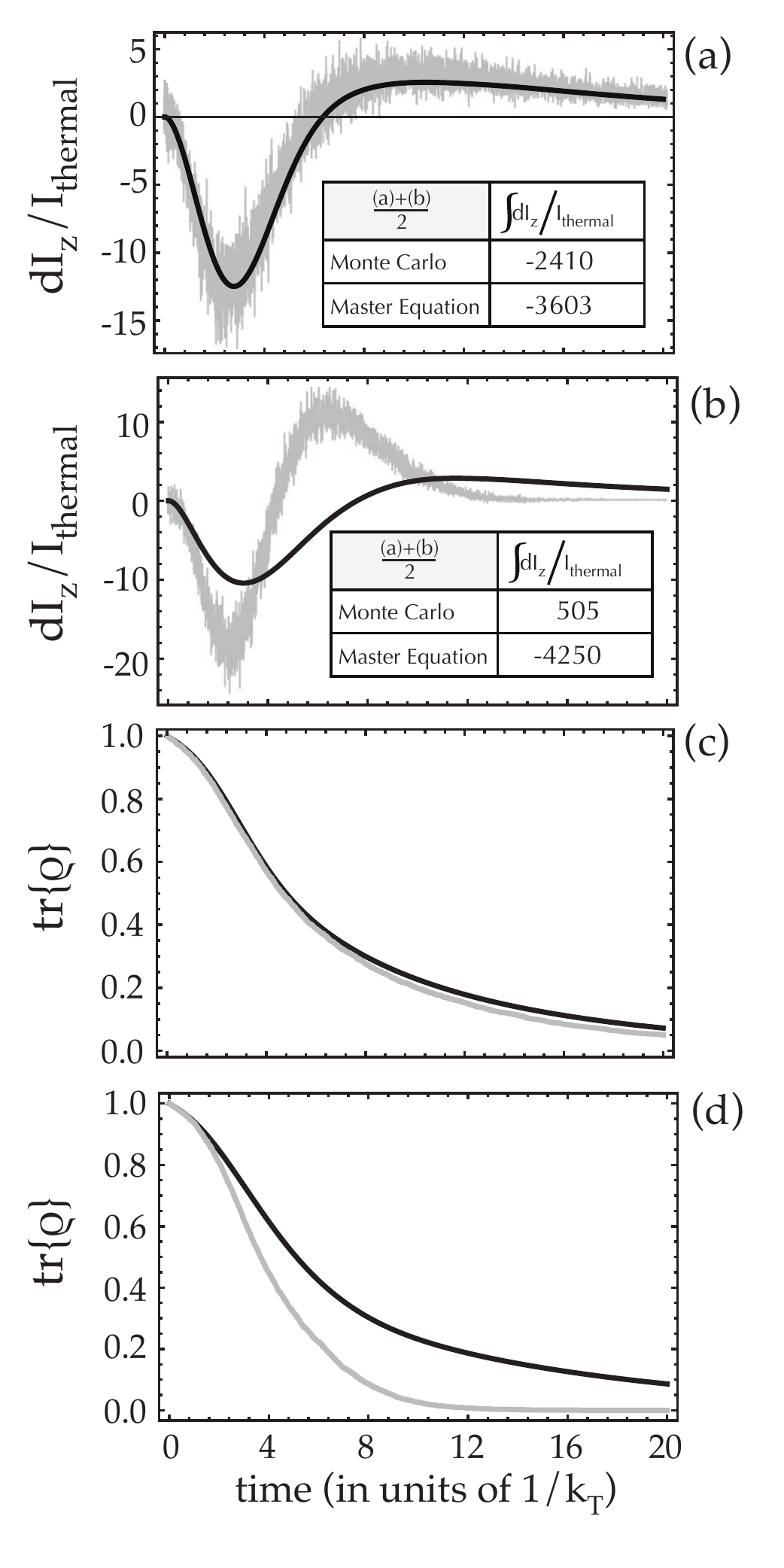}
\caption{Monte Carlo (grey) and master equation (black) calculation of (upper plots) the nuclear spin $dI_z$ deposited to the neutral reaction products normalized by $I_{\rm thermal}$, and (lower plots) the population of the RP ensemble, $\tr\{\rho\}$, as a function of time. (a,c) Kominis' and (b,d) Haberkorn's approach. In the insets of (a) and (b) we show the integral of the respective traces, reflecting the total nuclear spin of the reaction products at the end of the reaction. The Monte Carlo is the average of $2\times 10^5$ and $2\times 10^3$ quantum trajectories for (a,b) and (c,d), respectively, generated as explained in \cite{cpl2015}. Much fewer trajectories are required for simulating $\tr\{\rho\}$, since in each trajectory, $\tr\{\rho\}=1$ until the time of recombination and zero afterwards, while the nuclear spin deposited to the reaction products is zero everywhere except at the time of recombination.}
\label{dIz}
\end{center}
\end{figure}
We consider two different initial conditions, $|\psi_0\rangle=\ket{\rm S}\otimes\ket{\Downarrow}$ and $|\psi_0\rangle=\ket{\rm S}\otimes\ket{\Uparrow}$, the results of which we average in order to simulate the realistic scenario of starting with unpolarized nuclear spins. Finally, we integrate the averaged traces in order to find $\int dI_z$, which we normalize by $I_{\rm thermal}\approx 10^{-5}$, the Boltzmann equilibrium nuclear spin. The ratio $\int dI_z/I_{\rm thermal}$ is the so-called CIDNP enhancement, directly measurable in experiments. 

The results of the consistency check are shown in Fig. \ref{dIz}a and Fig. \ref{dIz}b (reproduced from \cite{cpl2015}). The level of inconsistency of Haberkorn's approach is evident by visually inspecting Fig. \ref{dIz}b. Even more impressive is the result for the enhancement $\int dI_z/I_{\rm thermal}$, shown in the insets of Fig. \ref{dIz}(a,b). The traditional theory is inconsistent both in the sign and the magnitude of the effect at the level of 740\%, whereas in our approach (Fig. \ref{dIz}a) the two results are of the same sign and different in magnitude by 50\%. 

As expected, some observables are not as sensitive to reaction kinetics as others. For example, in Fig. \ref{dIz}(c,d) we plot $\tr\{\rho\}$ as a function of time. Again, by qualitatively looking at Fig. \ref{dIz}d, Haberkorn's theory is immediately ruled out. However, at early times the inconsistency is rather innocuous in this case. For example, the reaction time (1/e of the initial population) read off the plot is about $6/\kt$ and $8/\kt$ for the Monte Carlo and the master equation, respectively, which are not alarmingly different. 

To summarize, using Haberkorn's master equation to extract physical parameters of photosynthetic reaction centers from CIDNP data will lead to significantly skewed numbers, since both the structure of the Hamiltonian and the relevant couplings will be forced by the incorrect reaction terms of \eqref{trad} to attain unphysical values in order to match the data. The level of the discrepancy can be gauged by comparing our Monte Carlo result for $\int dI_z/I_{\rm thermal}$ (-2410) with Haberkorn's master equation result (-4250). If the scientific community studying photosynthetic reaction centers is content with somewhat better than order-of-magnitude estimates of the relevant physical parameters, Haberkorn's approach is adequate. If precision is of interest, the conventional approach is inadequate.

Finally, our master equation succeeds in matching our Monte Carlo, which precisely captures the full quantum dynamics at the single molecule picture, at the level of 50\% for $\int dI_z$. This constitutes significant progress in establishing a fundamentally sound understanding of RP quantum dynamics, but this level of consistency is still not good enough. The weakness of our master equation is traced to the definitions of $p_{\rm coh}$ and $\rho_{\rm coh}$, which will be refined in forthcoming works.
\section{Competing theoretical approaches}
Soon after our first paper \cite{pre2009} challenging the traditional theory, yet another master equation based on quantum measurement considerations \cite{JH}, and a microscopic derivation of the traditional master equation \cite{ivanov} appeared in the literature. To these was recently added a study \cite{briegel2014} using the formalism of quantum maps. 
\subsection{The derivation of Jones \& Hore}
In \cite{JH} the authors use the operator sum approach to evolve the RP density matrix, encapsulating the physics of the atom being "watched" by an array of photomultipliers (Section 3.3), namely the assumption that a radical-pair that could recombine in e.g. the singlet channel, and was not observed to do so, is projected to the triplet state. This assumption derives from considering the RP spin degrees of freedom to be directly coupled to a physically observable environment, like the atom is coupled to the observable radiation field. 

\begin{figure}
\begin{center}
\includegraphics[width=8.5 cm]{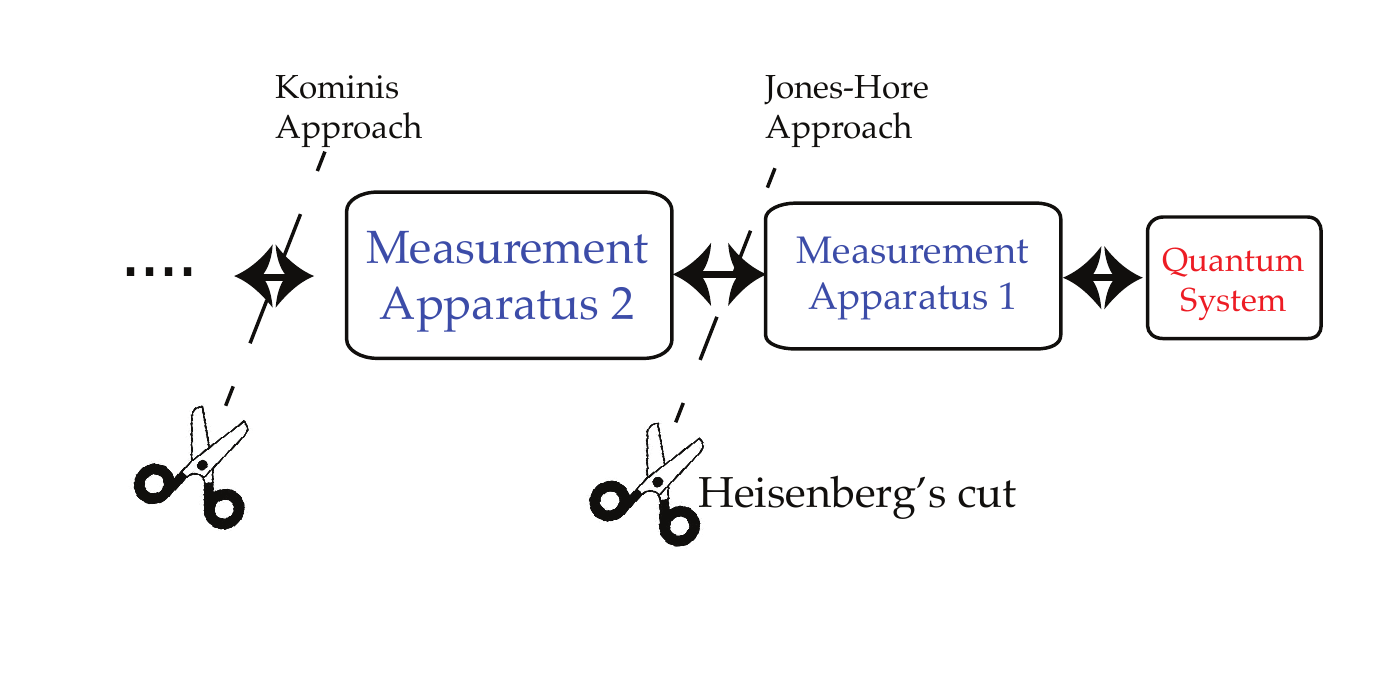}
\caption{The Jones-Hore approach assumes that RPs are directly "observed" by their recombination products. In our approach this is physically impossible because of the intermediate vibrational excitations of Fig. 5, which stand between the RP spin degrees of freedom and the neutral ground states. Thus the two approaches cut Heisenberg's measurement chain at different places.}
\label{cut}
\end{center}
\end{figure}

Put differently, considering Heisenberg's measurement chain depicted in Fig. \ref{cut}, where a quantum system interacts with an apparatus, which interacts with another apparatus and so on and forth until registering an actual measurement, the approach of Jones \& Hore is tantamount to cutting the chain right after the first apparatus. According to our understanding this is physically impossible, because the first apparatus is realized by the unobservable vibrational excitations. These are coupled to the observable phonon bath (second apparatus), so we can physically register the recombination event only with the detection of a phonon resulting from the decay of the vibrational excitations to the ground state. The absence of such a detection cannot unambiguously condition the radical-pair's state as in the atom watched by photodetectors, because of the unobservable possibility of virtual transitions to the vibrational reservoir and back to the RP.

Thus, in our approach, the presence of the intermediate "apparatus" between the radical-pair and the final product detection ameliorates the influence of the "null recombination-product measurement" on the RP's state. In the Jones-Hore approach, this influence is intensified due to the absence of the intermediate apparatus. This is why the Jones-Hore theory predicts a higher S-T dephasing rate than our theory. A recent experiment \cite{maeda2013} performed pulsed-EPR spectroscopy on a carotenoid-porphyrin-fullerene triad to find the S-T dephasing rate to be lower than the Jones-Hore prediction, ruling out the theory, since the theoretically expected fundamental relaxation rate always sets a lower bound to measured rates. The S-T dephasing rate measured in \cite{maeda2013} is consistent with Haberkorn's prediction, apart from a postulated extra relaxation mechanism attributed to g-anisotropy. The magnitude of this extra relaxation rate is roughly estimated in \cite{maeda2013} to be factor of 2 or 3 (depending on temperature) higher than Haberkorn's dephasing rate. Even if this attribution turns out to be correct, it is already understood that in the presence of strong hyperfine relaxation, Haberkorn's theory is expected to be a good approximation. Hyperfine relaxation is for S-T mixing in radical-pairs what Doppler broadening is for atoms. In order to study natural broadening of an atomic transition, Doppler broadening must be eliminated. In radical-pairs, in order to study the fundamental S-T dephasing we have introduced, one needs to experiment with molecules not suffering hyperfine relaxation {\it within their lifetime}. The molecule used in \cite{maeda2013} is not one of those. 

In contrast, another experiment \cite{molin2013,molin2013_2} reported evidence of a non-trivial state evolution of non-recombining radical-pairs, the rate of which appears to be consistent with the Jones-Hore prediction. A global and more detailed analysis of these two conflicting experiments will be undertaken elsewhere. 
\subsection{The derivation of Ivanov and co-workers}
The microscopic derivation of Ivanov et al. in \cite{ivanov} seamlessly arrives at Habekorn's master equation. The problem is that this derivation describes a system other than radical-pairs. This is because the authors consider the neutral reaction products to be {\it coherently} coupled to the radical pairs. However, from the electron-transfer picture of Fig. 5 and Fig. 6, it is clear that the presence of the continuous manifold of vibrational excitations of the neutral products prohibits any coherence between reactants and products, as long as by "product" we identify the physically observable neutral ground state. So is the master equation a matter of semantics, i.e. defining what is the reaction product? Not really. Again, it is the physics of the virtual transitions to the intermediate states and back to the RP that do not allow one to proclaim the reaction is over just counting on a non-zero amplitude of the intermediate states.

\begin{figure}
\begin{center}
\includegraphics[width=8.5 cm]{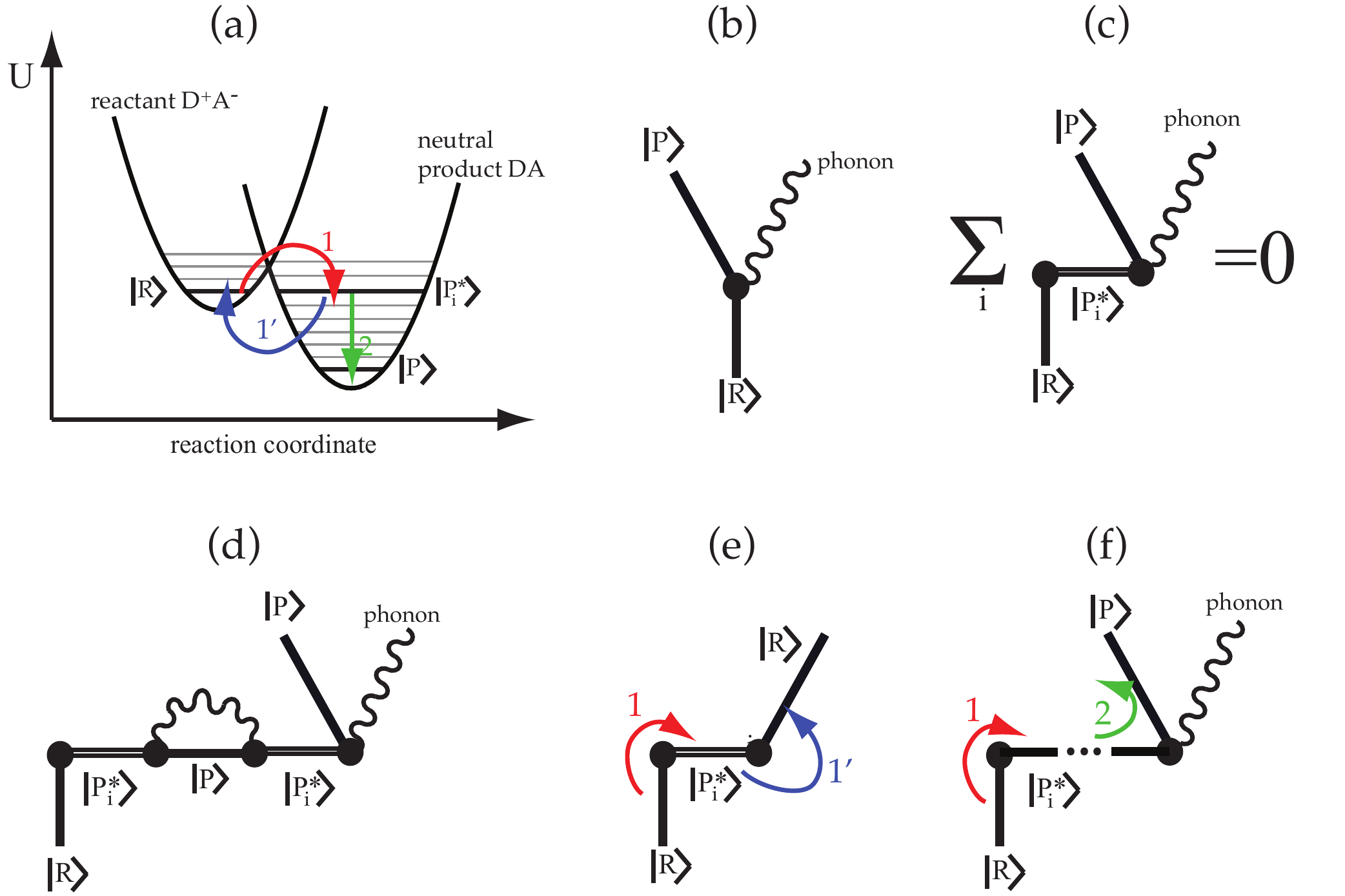}
\caption{(a) Energy level diagram of an electron-transfer reaction. (b) Effective reactant-product coupling neglecting intermediate states $|{\rm P}_{i}^{*}\rangle$. (c) Within 2$^{\rm nd}$-order perturbation theory the coherent reactant-product coupling interferes to zero. (d) Leading (4$^{\rm th}$) order coherent reactant-product coupling. (e) Reactant dephasing, resulting from a virtual transition to the intermediate states and back. (f) Product formation, resulting from a real transition from the reactant to the intermediate state and the decay of the latter to the product state. Essentially, (f) consists of two separate diagrams materializing consecutively, i.e. $|{\rm P}_{i}^{*}\rangle$ is here a real and not a virtual state like in (c-e).}
\label{RPcoh}
\end{center}
\end{figure}

These considerations have been analyzed in detail in \cite{pre2012} with the visualizing help of Feynman diagrams shown in Fig. \ref{RPcoh}. The coherent reactant-product coupling postulated \cite{ivanov} to describe electron-transfer reactions (Fig. \ref{RPcoh}a) is shown in Fig. \ref{RPcoh}b. This represents an "effective" theory neglecting the existence of the intermediate states, much like the early theory of nuclear $\beta$-decay neglected the existence of electroweak bosons. In reality, the intermediate states prevent the build-up of reactant-product quantum coherence, because the amplitudes for such a coupling summed over all intermediate states interfere away to zero (Fig. \ref{RPcoh}c). This is because of the energy denominator appearing in perturbation theory, which is an odd function of the energy difference between reactant and intermediate states \cite{pre2012}. The first non-zero amplitude for coherently coupling reactants with products comes from 4$^{\rm th}$-order perturbation theory (Fig. \ref{RPcoh}d) and is irrelevant for reaction kinetics. In Fig. \ref{RPcoh}e we show a diagram responsible for radical-pair S-T dephasing, which involves a virtual transition to the intermediate states (1) and back (1'). Finally, the product is formed by two consecutive real transitions, (1) from the reactant to the intermediate state, and (2) from the intermediate state to the product (Fig. \ref{RPcoh}f). To summarize, omitting intermediate states in irreversible reactions can lead to conceptual rigmaroles like states $|\psi\rangle=\Big(|\bomb\rangle+|\explosion\rangle\Big)/\sqrt{2}$, i.e. coherences between an exploded and un-exploded bomb. 

We can now make a point regarding our approach. Diagrams (e) and (f) of Fig. \ref{RPcoh} have different final states, hence do not interfere, and this is why to calculate ${{d\rho}\over {dt}}$ in our master equation \eqref{ME} we add ${{d\rho}\over {dt}}\big|_{\rm decoh}$, represented by Fig. \ref{RPcoh}e, with ${{d\rho}\over {dt}}\big|_{\rm recomb}$, represented by Fig. \ref{RPcoh}f.
\subsection{The derivations of Briegel and co-workers}
Briegel and co-workers considered the description of RP quantum dynamics from the perspective of quantum maps in \cite{briegelJPC} and more comprehensively in \cite{briegel2014}, where the authors attempt a generalized formulation of the problem, at the cost though, of not offering a clear-cut answer to the question an experimentalist, for example one detecting solid-state NMR signals from photosynthetic reaction centers, might rightfully ask: "is Haberkorn's approach adequate to account for the data and if not, which is the master equation one should use?". The answers given in \cite{briegel2014} are "maybe" and "it depends". We reply "no" and Eq. \eqref{ME}. 

In \cite{briegel2014} the authors invent an abstract environment, without specifying where this environment comes from and what are the microscopic physics coupling the RP's spin degrees of freedom to this environment. The authors then go on to provide several master equations given in terms of decay and dephasing amplitudes, which however, the "user" is supposed to specify. For example, given the typical scenario of a solid-state CIDNP experiment, i.e. an immobilized RP characterized by the two rates $\ks$ and $\kt$, and in the absence of any technical decoherence, it is not clear in \cite{briegel2014} what is the master equation one ought to use.

We stress the physical scenario of immobilized radical-pairs, because the authors in \cite{briegel2014} elevate molecular re-encounters to a fundamental physical effect, something that we think is more perplexing than illuminating. This is because the recombination rates $\ks$ and $\kt$ are in principle known functions of the distance between the two radicals. So if one has solved the immobilized RP case, one can readily generalize to a liquid state scenario, where the two radicals diffuse and re-encounter, and where $\ks$ and $\kt$ would become functions of time through their dependence on inter-radical separation. To summarize, the considerations of \cite{briegel2014} should become more precise in order that (i) it is discernible if and where they differ from other approaches, (ii) they are  amenable to criticism, and (iii) they are useful to experimentalists. 
\section{The physics of the nonlinear master equation}
The nonlinear nature of our master equation \eqref{ME} is a subtle point requiring a thorougher examination than we can afford here. Although we sympathize with the unease of many  quantum physicists with nonlinear master equations, we understand the nonlinearity of \eqref{ME} to be a necessary evil forced upon us by the physics of the problem. We will elucidate this point by a few thought experiments. In all of them we consider no S-T mixing (${\cal H}=0$) and an unreactive triplet state ($\kt=0$).

Suppose that we prepare two RP ensembles, one in the maximally coherent state $|\psi_1\rangle=(|{\rm S}\rangle+|{\rm T}_0\rangle)/\sqrt{2}$, or equivalently $\rho_1=|\psi_1\rangle\langle\psi_1|$, and the other in the maximally incoherent mixture of singlet and triplet states, $\rho_2={1\over 2}\ket{\rm S}\bra{\rm S}+{1\over 2}\ket{\rm T_0}\bra{\rm T_0}$. In Section 3.9.1 we presented the counterintuitive result that in the first ensemble we end up with 25\% of the original RPs locked in the non-reactive triplet state, while in the second we will end up with 50\% of the original RPs locked in the non-reactive triplet state. Both ensembles have the same $\rho_{\rm SS}+\rho_{\rm TT}$, but different $\rho_{\rm ST}+\rho_{\rm TS}$. Hence the reaction depends on $\rho_{\rm ST}+\rho_{\rm TS}$. That this dependence is nonlinear is seen by considering two different ensembles that will occupy the rest of the discussion.

Consider the ensemble E$_1$ prepared in the state $|\psi_1\rangle=(|{\rm S}\rangle+|{\rm T}_0\rangle)/\sqrt{2}$ and E$_2$ in the state $|\psi_2\rangle=(|{\rm S}\rangle-|{\rm T}_0\rangle)/\sqrt{2}$. The two ensembles just differ in the sign of $\rho_{\rm ST}+\rho_{\rm TS}$. The end result is the same for both, that is, at $t\gg 1/\ks$  we are left with 25\% of the original RPs locked in the non-reactive triplet state. It is thus clear that the reaction {\it depends on the "absolute value" of S-T coherence, which is a non-linear operation}.

Now consider mixing the initial contents of E$_1$ and E$_2$ and offering the box E$_{1+2}$ to a good experimenter who can quickly, before any recombination takes place, establish that the density matrix of the system is $\rho_{1+2}=\big(|{\rm S}\rangle\langle{\rm S}|+|{\rm T}_0\rangle\langle{\rm T}_0|\big)/2$. Indeed, in spin chemistry experiments individual RPs are not accessible. At the highest laboratory magnetic fields, the typical (e.g. for an EPR transition) frequency is on the order of 10$^{11}$ Hz, the corresponding wavelength and resolvable volume being 1 mm and 1 mm$^3$, respectively. This volume contains a macroscopic number of RPs. Hence any spin-state measurement on the box E$_{1+2}$ (assumingly fast and non-destructive) will be a global measurement on all RPs inside the box, and it will be consistent with the maximally incoherent density matrix $\rho_{1+2}$. What will be the final population of non-reactive triplet RPs that will remain in the box? If we know the specific preparation, we conclude it will be 25\%, like it was for E$_1$ and E$_2$ individually. If we don't, like the experimenter who is offered the box E$_{1+2}$, we conclude it will be 50\%, as expected from $\rho_{1+2}$. 

Contradiction? Not quite. The issue at hand has to do with {\it proper} versus {\it improper} mixtures \cite{petroni,brown,masillo}. The case in which the experimenter is unaware of the specific state preparation is referred to as a {\it proper} mixture. There is maximum S-T coherence at the single molecule level both in E$_1$ and in E$_2$, but due to the mixing the whole box E$_{1+2}$ appears maximally incoherent. In contrast, if there is a genuine decoherence process, like S-T dephasing, leading to the same mixed state, one talks of an {\it improper} mixture. In this case there is no coherence at the single molecule level, and the box E$_{1+2}$ indeed contains a mixture of singlet and triplet RPs. 

There is an illuminating analogy with Young's double slit experiment \cite{pre2011}. Suppose that a light source emits perfectly coherent light and we register photons on the observation screen for a time $\tau$. An interference pattern will be formed. Suppose then that we keep registering photons for a consecutive time interval $\tau$, this time putting a $\pi$ phase shifter in front one of the two slits. If we were to observe the two runs individually, we would see two perfect interference patterns shifted with respect to each other. On the other hand, if we wait to register all photons from the two consecutive runs, and then look at the observation screen, there will be no interference pattern. If one is presented with the final picture, one will conclude that the light source was incoherent. This is the analogy of the proper mixture discussed before. In contrast, if one performs this experiment, again with perfectly coherent light, and installs a measurement apparatus after the slits acquiring complete which-path information, the interference pattern will genuinely disappear. This is the analog of the improper mixture discussed above.

Going back to the bewildered experimenter who is provided the mixture $\rho_{1+2}$, while all his global measurements will lead him to expect 50\% unreacted triplet RPs, he will be surprised to finally measure 25\%. To our understanding, so far proper and improper mixtures were more of an epistemological curiosity rather than of practical utility. That is, we are unaware of any quantum science experiments able to differentiate proper from improper mixtures. We will close this review with a conjecture, the proof of which we leave as a problem to be addressed in the more abstract context of the foundations of quantum physics. {\it Leaky quantum systems, in which the leakage depends non-linearly on the quantum-state, can differentiate between proper and improper mixtures}.
\section{Outlook}
We hope to have convinced the reader that radical-pair reactions indeed constitute an ideal paradigm for quantum biology, since they require for their understanding the whole conceptual toolset of quantum information science, even touching upon the foundations of quantum physics. The complex quantum dynamics of radical-pair reactions emerge from the interplay of coherent spin dynamics and incoherent electron transfer reactions, rendering this system both open and leaky. To our knowledge, this is rather unusual in quantum science, and conceptually quite challenging, explaining {\it in part} the current lack of consensus in the scientific community on the quantum foundations of RPM. 

It is true that the last several years have witnessed several exciting discoveries making the case of quantum biology. We feel that the synthesis of the two seemingly disjoint fields that dominated 20$^{\rm th}$ century science, "quantum" and "bio", will lead to breakthroughs beyond our current ability to forecast. Hence we will be more than content if this review inspires many other researchers to join this tremendously exciting scientific endeavor. 

\section{Acknowledgments}

We acknowledge support from the European Union's Seventh Framework Program FP7-REGPOT-2012-2013-1 under grant agreement 316165.

\end{document}